\documentclass{aa}  
\usepackage{graphicx}
\usepackage{multicol}
\usepackage[para]{threeparttable}
\usepackage{txfonts}
\usepackage{placeins}
\usepackage{booktabs}
\usepackage[version=4]{mhchem}
\usepackage{amsmath,nccmath}
\usepackage{hyperref}
\usepackage{comment}
\newcommand{\asec}{$^{\prime\prime}$}

\newcommand{\nndp}{N$_2$D$^+$}
\newcommand{\nnhp}{N$_2$H$^+$}

\newcommand{\h}{H}
\newcommand{\hh}{H$_2$}

\newcommand{\hhhp}{H$_3^+$}
\newcommand{\hhdp}{H$_2$D$^+$}

\newcommand{\hcop}{HCO$^+$}

\def\SigmaH2{$\Sigma $(${\rm H_2}$)}

\def\H{N$_{2}$H$^{+}$}
\def\D{N$_{2}$D$^{+}$}
\def\15N{$^{15}$NNH$^+$}
\def\N15{N$^{15}$NH$^+$}

\def\h3p{H$_3^+$}
\def\h2dp{H$_2$D$^+$}

\def\AMM{NH$_3$}

\def\CII{\mbox{C$^{18}$O}}
\def\CIII{\mbox{C$^{17}$O}}

\def\METH{CH$_3$OH}

\def\H13CN{\mbox{H$^{13}$CN}}

\def\kms{\mbox{km~s$^{-1}$}}
\def\cmc{cm$^{-3}$}
\def\cmq{cm$^{-2}$}

\def\solm{\mbox{M$_\odot$}}

\def\NH{$N$(H$_2$)}
\def\Tex{\mbox{$T_{\rm ex}$}}

\def\Tk{\mbox{$T_{\rm k}$}}

\def\kms{km\,s$^{-1}$}

\def\hh{H$_2$}

\def\HMSC{G034}
\def\HMPO{AFGL}
\def\UCHII{G589}

\begin{document} 

\title{CHemical Evolution in MassIve star-forming COres (CHEMICO)}

\subtitle{II. Molecular ions, electron fraction, and cosmic-ray ionisation rate}

   \author{F. Fontani
          \inst{1,2,3}
        \and
        E. Roueff\inst{2}   
        \and
        V.M. Rivilla\inst{4}
        \and
        \'A. L\'opez-Gallifa\inst{4}
        \and
        L. Bizzocchi\inst{5}
        \and
        L. Colzi\inst{4}
        \and
        H. Mart\'in-Caballero\inst{6}
        \and
        P. Caselli\inst{3}
          }

    \institute{INAF - Osservatorio Astrofisico di Arcetri,
    Largo E. Fermi 5,
    I-50125, Florence, Italy \\
              \email{francesco.fontani@inaf.it}
         \and
         LUX, Observatoire de Paris, PSL Research University, CNRS, Sorbonne
         Universit\'e, F-92190 Meudon, France
         \and
         Max-Planck-Institut f\"{u}r extraterrestrische Physik, Giessenbachstra\ss e 1, 85748 Garching bei M\"{u}nchen, Germany
         \and
        Centro de Astrobiolog\'ia (CAB), CSIC-INTA, Ctra. de Ajalvir Km. 4, 28850, Torrej\'on de Ardoz, Madrid, Spain
        \and
        Dipartimento di Chimica “Giacomo Ciamician”, Universit\`a di Bologna, Via P. Gobetti 85, I-40129 Bologna, Italy
       \and
        Vrije Universiteit Amsterdam De Boelelaan 1105, 1081 HV Amsterdam, The Netherlands
             }

   \date{Received xxx; accepted xxx}

  \abstract
{Molecular ions are crucial in initiating and regulating the formation of molecules in the interstellar medium, and are expected to be sensitive tracers of evolution in star-forming regions, as they are linked to specific physical properties that change over time.}
{Within the framework of the project `CHemical Evolution of MassIve star-forming COres' (CHEMICO), we investigate the temporal evolution of molecular ions towards three high-mass star-forming cores that represent three sequential evolutionary stages.} 
{CHEMICO is based on an unbiased spectral line survey of the bandwidth at 3, 2, and 1.2~mm carried out with the 30m telescope of the Insitut de Radioastronomie millim\'etrique.}
{We detect 19 molecular ions, including isotopologues. In particular, we report the first tentative detection of the rare species PO$^+$, H$_2$COH$^+$, CH$_3$CO$^+$, and HC$_3$NH$^+$ towards the more evolved sources. 
From a local thermodynamic equilibrium analysis, we derive the following main results:
(i) Overall, the physical parameters of sulphur-bearing ions are the most sensitive to evolution, whereas those of nitrogen-bearing ions are the least;
(ii) The molecular abundances that exhibit the most pronounced temporal variations are those of SO$^+$, DCO$^+$, and \nndp; 
(iii) A simple analytic approach tailored for warm gas conditions provides electron fractions $\sim 10^{-8}-10^{-7}$ and cosmic-ray ionisation rates of H$_2$ $\sim 10^{-16}-10^{-13}$ s$^{-1}$, several orders of magnitude above the canonical values measured in the interstellar medium and with a marked increase with evolution;
(iv) The abundance ratio of protonated-to-neutral species increases with the proton affinity of the neutral regardless of the evolutionary stage of the source, indicating that proton-transfer appears to proceed similarly in both cold and warm interstellar environments.}
{Molecular ions exhibit distinct evolutionary trends in high-mass star-forming regions, with sulphur-bearing species and the abundances of \D , DCO$^+$, and SO$^+$ emerging as particularly sensitive tracers of evolution.
Measured electron fractions and cosmic-ray ionisation rates are in line with an internal production of cosmic-rays that becomes increasingly efficient with time.}


   \maketitle
\nolinenumbers

\section{Introduction}
\label{intro}

It is now clear that our Sun was born in a crowded
stellar cluster including high-mass stars, i.e. stars more massive than $\sim 8$~\solm\ (e.g. Adams~\citeyear{adams10},
Pfalzner et al.~\citeyear{pfalzner15}, Jensen et al.~\citeyear{jensen19}, Arzoumanian et al.~\citeyear{arzoumanian23})
, even though the Galactocentric distance of the place where the Sun was born is still debated \citep[e.g.,][]{baba24}.
The case of the Sun is common, 
as it is now well established that most stars are born in rich clusters (e.g.~Carpenter~\citeyear{carpenter00}; 
Lada \& Lada~\citeyear{lel03}) which included high-mass stars (e.g.~Rivilla et al.~\citeyear{rivilla13},~\citeyear{rivilla14}).
Despite this, our knowledge of how high-mass stars 
are born is still limited, owing to both observational and theoretical problems 
(e.g.~ Tan et al.~\citeyear{tan14}; Krumholz~\citeyear{krumholz15}; Motte et al.~\citeyear{motte18}; Padoan et al.~\citeyear{padoan20}). 
Observations are challenging because high-mass star-forming cores (i.e. sub-pc scale structures where high-mass stars are born) are fewer.
Moreover, being located at larger distances, they have smaller angular dimensions with respect to local low-mass ones.
Also, they are typically affected by large amounts of ambient gas and/or nearby star formation activity. 
Theoretical problems arise primarily from the fact that high-mass stars reach the main sequence while still accreting, and the radiation pressure of the embryo star  
could be sufficient to impede any further spherical accretion \citep{les71,aet74}. 
Models can solve this problem invoking different initial conditions and accretion mechanisms for a high-mass star to form: either from single, isolated, high-mass pressure-confined cores where gas funnels inward through a circumstellar disk \citep[e.g.][]{met03}, or from low-mass `seeds' in a rich protocluster fed by a much larger parsec-scale molecular clump reservoir, drawing the bulk of its mass from well outside the immediate core environment \citep[e.g.][]{bonnell04}.
However, which of these models better matches observations remains debated, and the evolutionary phases are not separated as clearly 
as is the case for the low-mass star formation process (see,  e.g.,  Shu,  Adams  \&  Lizano~\citeyear{shu87};  Andr\'e,  Ward-Thompson \& Barsony~\citeyear{andre00}). 

\begin{figure}
   \centering
   \includegraphics[width=9cm]{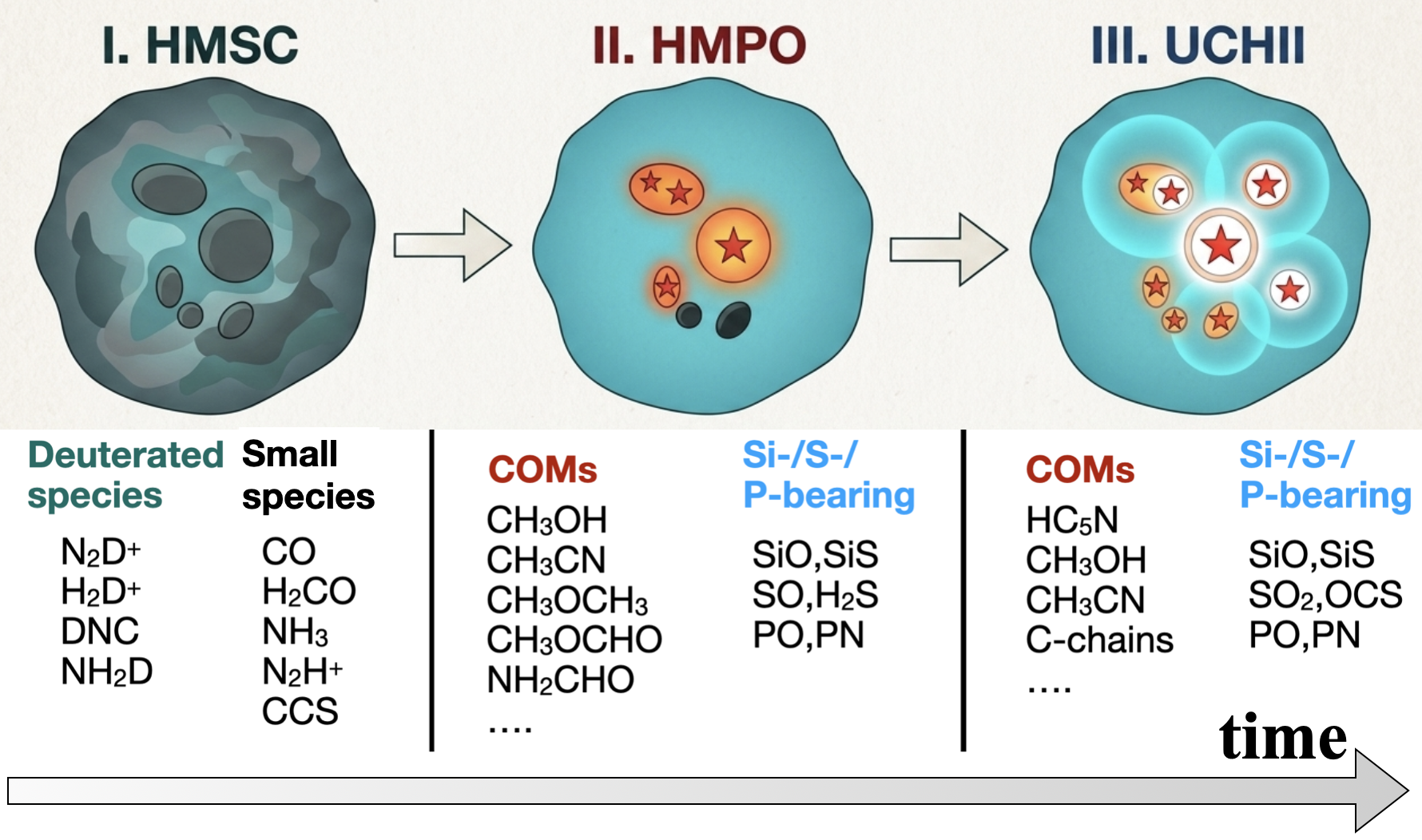}
      \caption{Coarse evolutionary classification for high-mass star-forming regions in high-mass starless cores (HMSCs), high-mass protostellar objects (HMPOs), and ultracompact HII regions (UCHIIs).
      The main chemical indicators of each evolutionary group are listed.
      Small species include those with less than 5 atoms, while COMs represent organic species with 6 or more atoms.
      Adapted from \citet[][]{fontani25a}.
              }
    \label{fig:sequence}
\end{figure}

Several attempts have been made to find an empirical temporal classification of pc-scale high-mass star-forming cores (e.g.~Tan et al.~\citeyear{tan14}; Motte et al.~\citeyear{motte18}; Beuther et al.~\citeyear{beuther25}). 
Following the sequence established for low-mass star-forming cores, 
these schemes typically divide the evolution into three main phases shown in Fig.~\ref{fig:sequence}.
The earliest stage is that of high-mass starless cores (HMSCs), infrared-dark molecular clouds on the verge of gravitational collapse characterised by low kinetic temperatures (\Tk $\sim 10-20$~K) and high H$_2$ volume densities ($n\geq 10^{3}-10^{5}$~\cmc, depending on size). 
When the gravitational collapse starts, these cores transition into high-mass protostellar objects (HMPOs), infared-bright cores characterised by densities and temperatures higher than in the previous stage ($n\simeq 10^6$~\cmc, \Tk $\geq 20$~K).
Finally, when (some of) the embedded protostars become zero-age main sequence stars, the HMPOs transition to ultra-compact HII regions (UCHIIs), characterised by expanding ionised bubbles produced by the newly born stars surrounded by a neutral molecular cocoon ($n\geq 10^5$~\cmc, 
\Tk $\sim 20 - 100$~K) influenced by the progressive expansion of the ionised bubbles.

Such physical evolution, associated with significant variations in temperature and density, is expected to proceed in parallel with a chemical evolution.
Previous studies have highlighted potential evolutionary indicators of specific phases.
For example, as shown in Fig.~\ref{fig:sequence}, deuterated molecules can be used to study the earliest stages \citep[e.g.][]{fontani11,fontani15a,sabatini24}, while complex organic molecules (COMs) and sulphur-bearing species are suitable for later phases \citep[e.g.][]{coletta20,fontani23}.
However, identifying good molecular links from the early to the late stages remains a significant challenge \citep[][]{fontani25a}.
A major issue lies in the fact that time is just one of the ingredients regulating the chemical evolution.
Very often, source surveys contain objects considered almost coeval but with different physical and environmental properties (in particular density, temperature, external irradiation, local metallicity), making it difficult to isolate the role of time in determining molecular abundances (see the discussion in Fontani et al.~\citeyear{fontani25a}).
In this context, observations of sources with well-constrained physical properties and reliably determined relative evolutionary stages are particularly valuable, as they help disentangle the effects of time from those of the environmental conditions and thereby reduce the degeneracy in the interpretation of molecular abundances.


The project `CHemical Evolution in Massive star-forming COres' \citep[CHEMICO,][hereafter paper I]{fontani25b} presents the first full spectral survey in the millimetre bands with the Institut de RadioAstronomie Millim\'etrique (IRAM) 30m telescope
towards three dense cores that best represent the three evolutionary stages of high-mass star formation. 
The main goal of CHEMICO is to derive column densities and fractional abundances of all the molecules detected from all observable transitions, and inspect relations with the evolutionary stage.
In paper I we presented the dataset, the source sample, the methodology, and the analysis of some carbon-chain and hydrocarbon species, from which we derived the evolution of the temperature structure, essential to properly interpret the chemistry.
In this work we analyse the molecular ions.

So far, more than 40 molecular ions have been identified in the interstellar medium\footnote{\url{https://cdms.astro.uni-koeln.de/classic/molecules}}.
In dense star-forming regions, where the gas is mostly neutral, ions can be created by cosmic-rays (CRs).
The process starts by the ionisation of H$_2$ through CRs with energy smaller than GeV to form H$_2^+$ \citep{grenier15}.
Once formed, H$_2^+$ rapidly reacts with neutral H$_2$ to produce H$_3^+$.
H$_3^+$ is a pivotal molecule that initiates the gas-phase chemistry of carbon- and oxygen-bearing molecules thanks to its low proton affinity \citep[e.g.,][]{herbst00}.
For example, the reaction of H$_3^+$ with atomic C leads, eventually, to the production of simple hydrocarbons, while the reaction with atomic O leads to the production of water \citep[e.g.,][]{watson73}.
Molecular ions regulate vast reaction networks in a variety of star formation environments \citep[e.g.,][]{hek73,herbst05,aikawa15,jorgensen20} mostly because ion-neutral reactions typically do not possess activation energy \citep[e.g.,][]{hek73}, and dominate traditionally the gas-phase chemistry at low temperatures.
Their abundances are tightly dependent on density, elemental abundances, CR ionisation rate, and temperature \citep[e.g.,][]{caselli02,vandertak05,podio14}.  
Because of this, molecular ions act as sensitive tracers of the physical conditions that shape the birth of stars and planets \citep[e.g.,][]{vandishoeck06,fontani21,luo24}.
Also, the ionisation carriers may involve atomic ions that drive a different chemistry and may contribute to a complex dynamical behaviour in interstellar chemistry  \citep{lebourlot93,lebourlot95,lee98}.

This work presents the analysis of all molecular ions observed in the three CHEMICO sources, and investigates which species are most sensitive to evolution. 
A link with the temporal evolution of the electron fraction and of the cosmic-ray ionisation rate of H$_2$ is also discussed.
The paper is organised as follows.
Section~\ref{sampleandobs} describes the source sample, the observations, and the methodology adopted for data reduction and analysis. 
Section~\ref{res} presents the observational results derived from the detected molecular ions, which we discuss in Sect.~\ref{discu}. 
Section~\ref{conc} summarises of the main findings of this paper, and discusses future perspectives.


\section{Sample, observations, and data analysis}
\label{sampleandobs}

The sources are extracted from the sample described by \citep{fontani11}, who studied 27 high-mass star-forming cores divided into the three evolutionary classes described in Sect.~\ref{intro}: 11 HMSCs, nine HMPOs, and seven UCHIIs. 
The CHEMICO targets are G034G2(MM2), hereafter G034, as representative
of HMSCs, AFGL5142--MM (hereafter AFGL) as the HMPO, and G5.89-0.39 (hereafter G589) as the UCHII.
The selection criteria are described in paper I. 
The interferometric continuum images are published by \citet{tan13} for \HMSC, \citet{rivilla20b} for \HMPO,
and \citet{hernandez14} for \UCHII. 
G034 contains two millimetre cores separated by $\sim 10$\asec, resolved in Atacama Large Millimeter Array (ALMA) images \citep{tan13}, but both are starless and thus the source reliably represents the first stage of the evolutionary sequence shown in Fig.~\ref{fig:sequence}.
Source coordinates and local standard of rest (LSR) velocities used to centre the receivers are listed in Table~\ref{tab:sources}. 

\begin{table*}
\caption[]{Source coordinates and basic physical parameters.}
    \label{tab:sources}
         \begin{tabular}{lccccccc}
            \hline
            \noalign{\smallskip}
            Source      &  R.A.\tablefootmark{a} & Dec.\tablefootmark{a} & V$_{\rm LSR}$\tablefootmark{b} & $d$\tablefootmark{c} & $L$\tablefootmark{c} & $N({\rm H_2})$\tablefootmark{d} & \Tk \tablefootmark{e} \\
                &  ($h:m:s$) & ($\circ:\prime:\prime\prime$) & (\kms) & (kpc) & (L$_{\odot}$) & (\cmq) & (K) \\
            \noalign{\smallskip}
            \hline
            \noalign{\smallskip}
        HMSC -- G034G2(MM2) [G034] & 18:56:50.0 & 01:23:08  &  +43.6   & 2.9 & $10^{1.6}$ & 4.1$\times 10^{22}$ & 15.2$\pm$0.4 \\
        HMPO -- AFGL5142--MM [AFGL] & 05:30:48.0 & 33:47:54 &  $-3.9$ & 1.8 & $10^{3.6}$ & 1.0$\times 10^{23}$ & 21.5$\pm$0.5 \\
        UCHII -- G5.89--0.39 [G589] & 18:00:30.5 & $-24$:04:01 & +9.0 & 1.3 & $10^{4.4}$ & 5.5$\times 10^{23}$ & 29.0$\pm$0.3 \\
            \noalign{\smallskip}
            \hline
         \end{tabular}
    \tablefoot{\tablefoottext{a}{J2000;}
    \tablefoottext{b}{Derived from either observations of $^{13}$CO (G034) or \AMM\ inversion transitions (AFGL, G589). Reference works are in \citet{fontani11};}
    \tablefoottext{c}{Reference works for distances are: \citet{bet09} for G034, \citet{snell88} for AFGL, and \citet{motogi11} for G589.
    Reference works for luminosities are: \citet{rathborne10} for G034, \citet{carpenter90} for AFGL, and \citet{mininni21} for G589};
    \tablefoottext{d}{From \citet{fontani18}, computed from the dust thermal continuum emission images at 850~$\mu$m obtained with the Submillimeter Common-User Bolometer Array (SCUBA) at the James Clerk Maxwell Telescope (JCMT) by \citet{difrancesco08}.}
    \tablefoottext{e}{derived from ammonia by \citet{fontani15a}.}
    }
   \end{table*}
 
Observations were conducted with the IRAM 30m telescope during two observing runs in springs 2021 and 2022 (projects 124--20 and 129--21, respectively). 
We covered almost the entirety of the three bands at 3, 2, and 1.3~mm of the Eight MIxer Receiver (EMIR) receiver (hereafter E0, E1 and E2) for the three targets, using the permitted combinations E0/E2 and E0/E1.
Band E3 was observed only towards \HMPO\ using the E1/E3 combination in the first observing run, and not towards \HMSC\ and \UCHII\ for the reasons described in paper I. 
Table~\ref{tab:obs} presents the observed spectral ranges, the beam full width at half maximum (HPBW), and the velocity resolution ($V_{\rm res}$).
The frequency resolution was 200 kHz in all bands.
Further information on the observations (calibration, background subtraction, pointing, focus, conversion from antenna temperature to main beam temperature scale) is given in paper I, where further details can be found.
Due to variable observational (mainly weather) conditions, the final root mean square (rms) noises are inhomogeneous. 
A representative range of achieved $1\sigma$~rms values is provided in Table~\ref{tab:obs}.

\begin{table}
    \caption[]{Observational parameters.}
    \label{tab:obs}
    \setlength{\tabcolsep}{2.5pt}
     \begin{tabular}{lcccc}
     \hline
        \noalign{\smallskip}
    source    &   bandwidth\tablefootmark{a}      &  HPBW\tablefootmark{b} & $V_{\rm res}$\tablefootmark{c} & $1\sigma$ rms\tablefootmark{d} \\
              &   (GHz)     & (\asec) & (\kms) & (mK) \\
            \noalign{\smallskip}
            \hline
            \noalign{\smallskip}
        \multicolumn{5}{c}{E0} \\
        \noalign{\smallskip} 
    G034 & 72.685--117.320 & $33.7-21$ & $\sim 0.82 - 0.51$ & 5--10 \\
    AFGL & 73--117 & $33.7-21$ & & 6--11 \\
    G589 & 72.685--115.255 & $33.7-21$ & & 6--12 \\
    \noalign{\smallskip}
    \hline
    \noalign{\smallskip}
     \multicolumn{5}{c}{E1} \\
    \noalign{\smallskip} 
    G034 &  124.685--182.205 & $19.7-13.4$ & $\sim 0.48 - 0.33$ & 7--14 \\
    AFGL &  125--184 & $19.7-13.4$ & & 11--15 \\
    G589 &  124.685--180.205 & $19.7-13.4$ & & 11--18 \\
    \noalign{\smallskip}
    \hline
    \noalign{\smallskip}
     \multicolumn{5}{c}{E2} \\
    \noalign{\smallskip} 
    G034 & 201.685--274.335 & $12.2-9.0$ & $\sim 0.30-0.22$  & 18--32 \\
    AFGL & 202.01--274.01 & $12.2-9.0$ & & 18--32 \\
    G589 & 201.685--274.335 & $12.2-9.0$ & & 18--34 \\
    \noalign{\smallskip}
    \hline
    \noalign{\smallskip} 
     \multicolumn{5}{c}{E3} \\
    \noalign{\smallskip} 
    AFGL & 284--347 & $8.7-7.1$ & $\sim 0.21 - 0.17$ & 45--85 \\
            \noalign{\smallskip}
            \hline
         \end{tabular}
         \tablefoot{\tablefoottext{a}{The E3 band was observed only towards AFGL;}\tablefoottext{b}{Half-power beam width of the telescope at the bandwidth frequency edges;}
         \tablefoottext{c}{Velocity resolution;}
         \tablefoottext{d}{In $T_{\rm MB}$ units.}
     }
\end{table}

Baseline-subtracted spectra in main beam temperature ($T_{\rm MB}$) units\footnote{The first steps of data reduction (e.g. averaging scans, baseline removal, conversion from antenna to main-beam temperature) were performed using the Continuum and Line Analysis Single-dish Software
({\sc class}) package of the Grenoble Image and Line Data Analysis Software ({\sc gildas}, \url{https://www.iram.fr/IRAMFR/GILDAS/}) following standard procedures.} were 
imported into the MAdrid Data CUBe Analysis ({\sc MADCUBA}\footnote{{\sc MADCUBA} is a software developed at the Madrid Centro of Astrobiolog\'ia (CAB, CSIC-INTA) for visualisation and analysis of single spectra and data cubes: \url{https://cab.inta-csic.es/madcuba/}.}, Mart\'in et al.~\citeyear{martin19}) software.
As in paper I, line identification and fitting were performed using the Spectral Line Identification and Modelling ({\sc slim}) tool of {\sc madcuba}, which includes, among others, the Cologne Database for Molecular Spectroscopy (CDMS\footnote{\url{https://cdms.astro.uni-koeln.de/cdms/portal/}}; \citealt{endres16}) and the Jet Propulsion Laboratory \citep[JPL,][]{pickett98} molecular spectroscopy database.
We used {\sc slim} to generate synthetic spectra of the targeted molecules under the assumption of local thermodynamic equilibrium (LTE) and compared with the observed spectra. 
Once a molecule was identified, we fitted the spectra using the \textsc{slim-autofit} tool, which provides the best non-linear least-squares LTE fit to the data using the Levenberg-Marquardt algorithm (more details in Mart\'in et al.~\citeyear{martin19}).
The parameters used in the LTE model were the molecular column density ($N$), excitation temperature ($T_{\rm ex}$), peak velocity ($V_\text{p}$), and full width at half maximum (FWHM) of the Gaussian line profiles. 
Because the source size, a requirement for accurately calculating the beam dilution factor (which could also vary for each transition), is unknown for the vast majority of the transitions, we assumed that the emission fills the telescope beam (i.e., no beam dilution).
This assumption is partially justified by the angular extension of the millimetre continuum emission ($\sim 10-20$\asec) as observed in high-angular-resolution images, and by the fact that the molecular emission is generally more extended (see reference works in Sect.~\ref{sampleandobs}).
However, since the transitions with high $E_{\rm up}$ (e.g. $\geq 100$~K) are expected to arise from compact regions that could be smaller than the telescope beam, the excitation temperatures derived for molecules including these transitions should be considered as lower limits.
The outcome of \textsc{slim-autofit} provides the value of the parameters, along with their associated uncertainties. 
Lines associated with a spectrally-resolved hyperfine structure were fit taking it into consideration.
For some species, \Tex\ had to be fixed, either because we did not detect a sufficient number of lines with different excitation conditions (at least three lines), or because the fit could not converge leaving \Tex\ free.
In Sect.~\ref{phys-par}, we will discuss case by case the reasons for the adopted \Tex.

Many spectral lines in G034 are best fitted with two velocity components separated in velocity by $\sim 1.8$~\kms, associated with the emission from the two cores detected with ALMA \citep[see above,][]{tan13} unresolved in the IRAM 30m telescope beam(s).
However, for NS$^+$, HCNH$^+$, and HOCO$^+$, the two velocity components are not spectrally resolved, and the best-fit is obtained with one component only.
All spectra of molecular ions detected towards AFGL are best fitted with a single velocity component.
Finally, the lines of H$^{13}$CO$^+$, HC$^{18}$O$^+$, and HCS$^+$ towards G589 spectra are best fitted with two velocity components (a narrower one and a wider one), due to the unresolved emission from the envelope and from the outflow, while all other species are better fitted with the narrow component only.

\section{Results}
\label{res}

\subsection{Detected molecular ions}
\label{detection}

\begin{table}
\caption[]{Molecular ions detected (Y), undetected (N), and tentatively detected (Y?).}
    \label{tab:detections}
    \setlength{\tabcolsep}{2.5pt}
         \begin{tabular}{lcccc|cc}
            \hline
            \noalign{\smallskip}
  Source             &  G034 & AFGL & G589 &  & \multicolumn{2}{c}{Undetected} \\
  Type                   &  HMSC & HMPO & UCHII & & Cations & Anions \\
            \noalign{\smallskip}
    \cline{1-4} \cline{6-7}
            \noalign{\smallskip}
          &      \multicolumn{3}{c}{Two atoms} & & CF$^+$ & CN$^-$ \\
                
NS$^+$   &  Y   &  Y  &   Y  & & CO$^+$ & C$_3$N$^-$ \\  
PO$^+$   &  N   &  Y?  &   N  & & HCl$^+$ & C$_4$H$^-$ \\
SO$^+$   &  Y   &  Y  &   Y  & & NO$^+$ & C$_5$N$^-$\\
\noalign{\smallskip}
           &     \multicolumn{3}{c}{Three atoms} & & D$^{13}$CO$^+$ & C$_6$H$^-$ \\
HCO$^+$     & Y  &   Y  &   Y & & DC$^{18}$O$^+$ & C$_7$N$^-$\\
DCO$^+$     & Y  &   Y  &   Y & & DCS$^+$ & C$_8$H$^-$ \\
H$^{13}$CO$^+$   & Y  &   Y  &   Y & & H$^{13}$CS$^+$ & \\
HC$^{18}$O$^+$   & Y  &   Y  &   Y & & H$_{2}$Cl$^+$ & \\
HC$^{17}$O$^+$   & N  &   Y  &   Y & & HOC$^+$ & \\
N$_2$H$^+$     & Y  &   Y  &   Y & & $^{15}$NND$^+$ & \\
N$_2$D$^+$     & Y  &   Y  &   Y & & N$^{15}$ND$^+$ & \\
$^{15}$NNH$^+$   & N  &   Y  &   Y & & C$_{3}$H$^+$ & \\
N$^{15}$NH$^+$   & N  &   Y  &   Y & & DOCO$^+$ & \\
HCS$^+$          & Y  &   Y  &   Y & & HCCS$^+$ & \\
HC$^{34}$S$^+$   & N  &   N  &   Y & & HOCS$^+$ & \\
\noalign{\smallskip}
            &     \multicolumn{3}{c}{Four or more atoms} & & HO$^{13}$CO$^+$ & \\
HCNH$^+$   &  Y  &   Y  &  Y & & C$_2$H$_3^+$ & \\
HOCO$^+$   &  Y  &   Y   &  Y & & HC$_3$O$^+$ & \\
H$_2$COH$^+$  &  N  &   Y?  &  Y? & & HC$_3$S$^+$ & \\
CH$_3$CO$^+$  &  N  &   N   &  Y?  & & H$_2$NCO$^+$ & \\
HC$_3$NH$^+$  &  N  &   N   &  Y? & & NCCNH$^+$ & \\
        &   &    &   & & HCCNCH$^+$ & \\ 
        &   &    &   & & H$_2$C$_3$H$^+$ & \\ 
        &   &    &   & & HC$_5$NH$^+$ & \\ 
        &   &    &   & & HC$_7$NH$^+$ & \\ 
            \noalign{\smallskip}
            \hline
         \end{tabular}
    \tablefoot{The last two columns contain species detected in the ISM not detected towards any of the three CHEMICO sources.
    }
   \end{table}


We searched for the emission of all molecular ions detected so far in the interstellar medium with transitions covered by the CHEMICO observations, following the methodology described in Sect.~\ref{sampleandobs}.
We considered as detected the species for which at least one transition has peak intensity $\geq 5\sigma$~rms, and the other transitions are consistent with non-detections according to the best-fit parameters.
We considered as tentatively detected the species for which at least one transition has peak intensity in the range $3-5\sigma$~rms, and the others eventually present in the spectrum are consistent with non-detections.
Table~\ref{tab:detections} reports the detected, tentatively detected, and undetected species.
Some species were already analysed in previous papers based on a lower number of lines.
These are \nnhp\ and its isotopologues \citep{fontani11,fontani15b}, HOCO$^+$ \citep{fontani18}, and HCNH$^+$ \citep{fontani21}.
Here, we have reanalysed these species with more lines. 
The best-fit results are consistent with those obtained in the previous works for HCNH$^+$ in all sources and for HOCO$^+$ towards G034, while for AFGL and G589 we derive HOCO$^+$ column densities a factor of $\sim 3$ and $\sim 2.3$ higher, respectively.
This difference is due to the different \Tex\ adopted.
In fact, when computing the column density using only the transition observed in \citet{fontani18}, and fixing \Tex\ to the one derived in this paper, we find column densities higher by a factor of 3.1 and 1.5  for AFGL and G589, respectively, than those given in \citet{fontani18}.
Considering the uncertainties, and the additional transitions included in the fit in our work, the values are consistent each other.
For \nnhp\ and isotopologues, we found a few significant discrepancies, which we discuss in detail in Sect.~\ref{n2dpanddcop}.

To corroborate some doubtful detections, in case the lines fell into spectral windows previously observed in other works \citep[in particular][]{fontani11,fontani15a,fontani15b}, we have merged the CHEMICO dataset with the previous ones to increase the signal-to-noise ratio.
This method allowed us to confirm and improve the detection of PO$^+$ $(2-1)$ and HCNH$^+$ $(3-2)$ towards \HMPO.
The spectroscopic parameters of the transitions detected in at least one target are listed in Table~\ref{tab:spectroscopic-tot}.
The spectra of all detected lines are shown in Appendix~\ref{app:A}.
As an example, in Fig.~\ref{fig:PO+} we show some selected transitions.

\begin{figure*}
    \centering
    \includegraphics[width=0.7\linewidth]{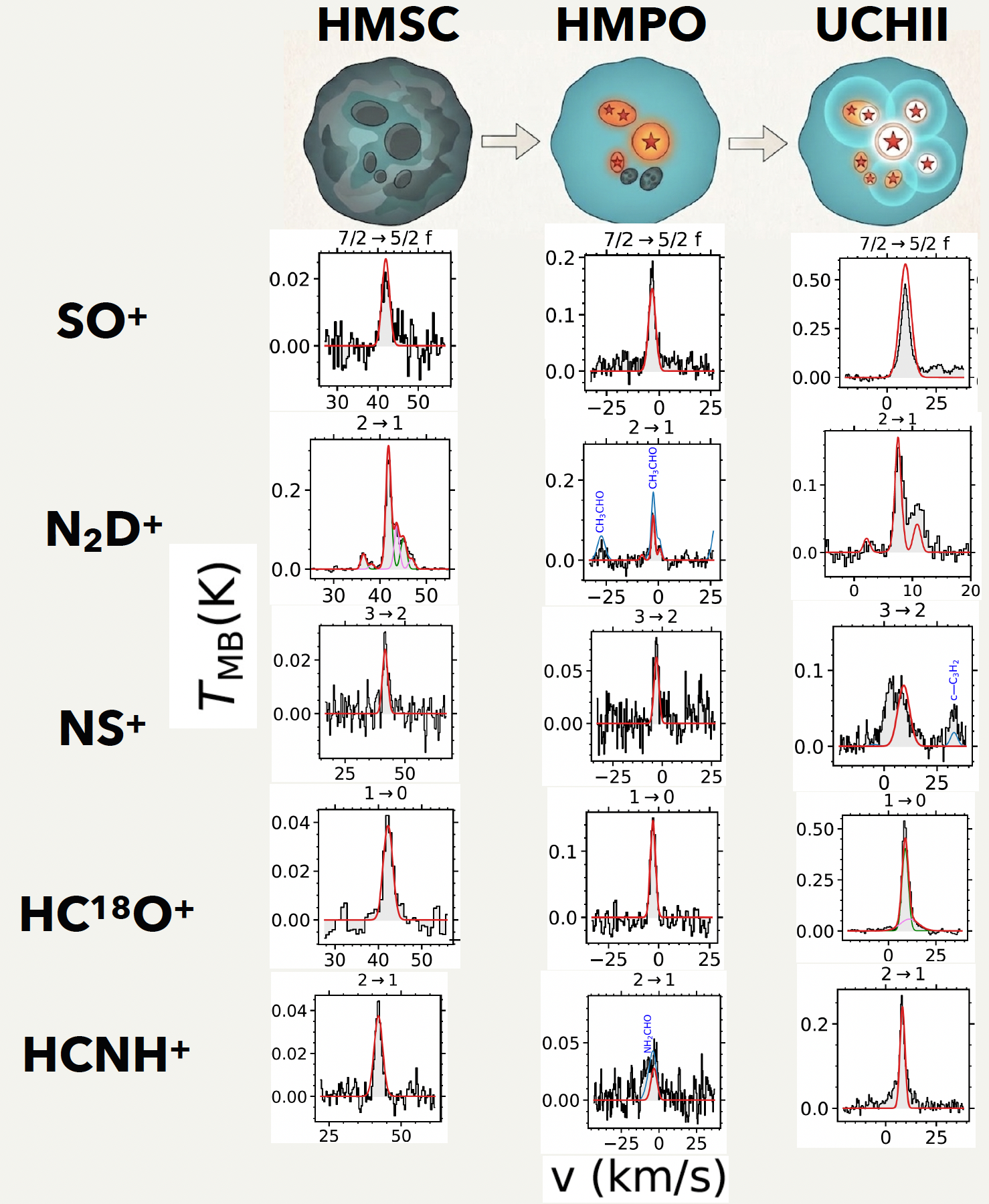}
    \caption{Sample spectra of detected molecular ions in each CHEMICO target. 
    The red curve indicates the best-fit obtained with {\sc madcuba}, which considers the line hyperfine structure for \nndp.
    The green and pink curves in \nndp\ for G034 and in HC$^{18}$O$^+$ for G589 correspond to the two velocity components
    (Table~\ref{tab:physical}), and the red curve in these spectra is the sum of these components. 
    The blue labels indicate nearby transitions of other species.}
    \label{fig:PO+}
\end{figure*}

\subsection{Best-fit line and physical parameters}
\label{phys-par}

\begin{figure}
   \centering
   \includegraphics[width=7.8cm]{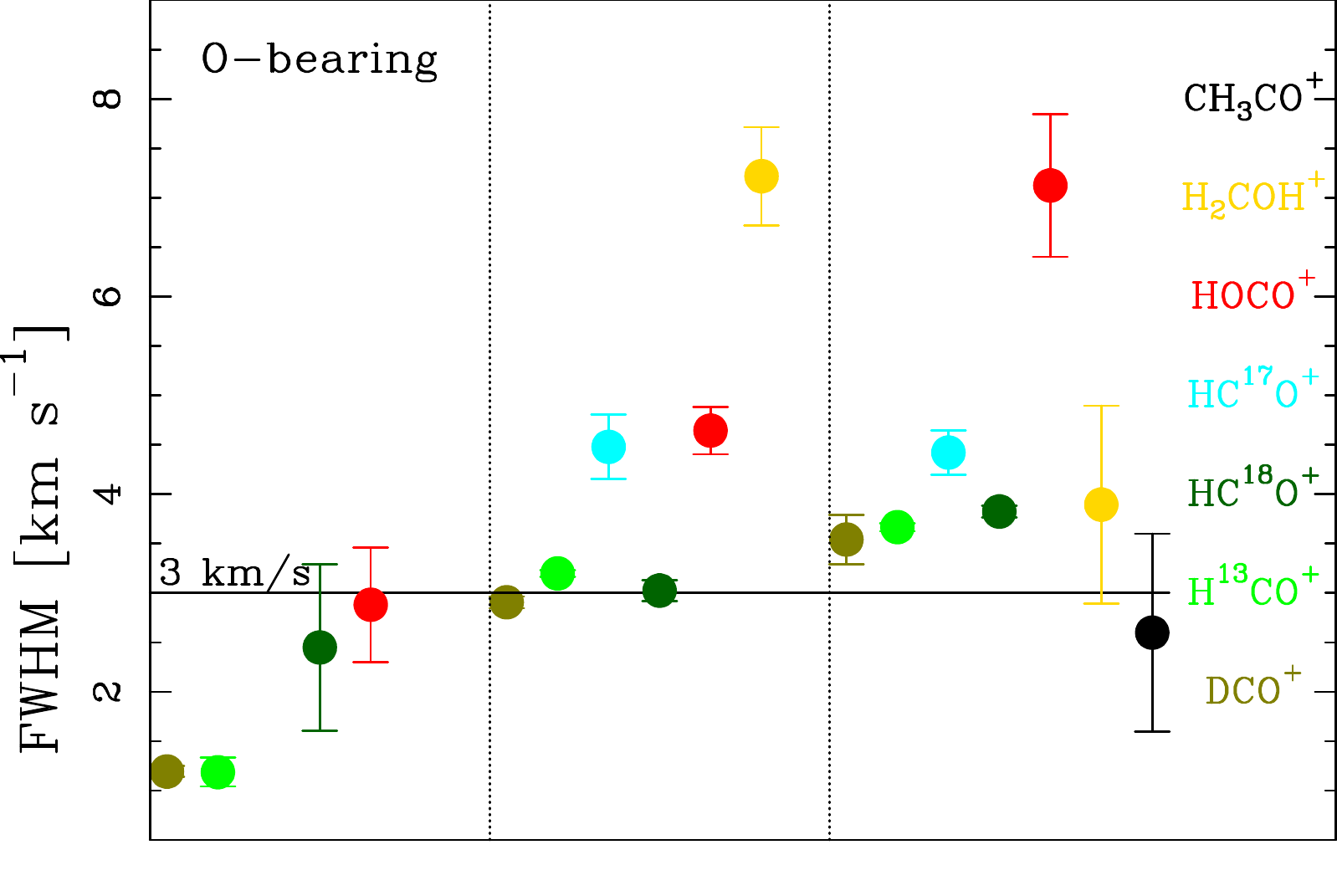}
   \includegraphics[width=7.8cm]{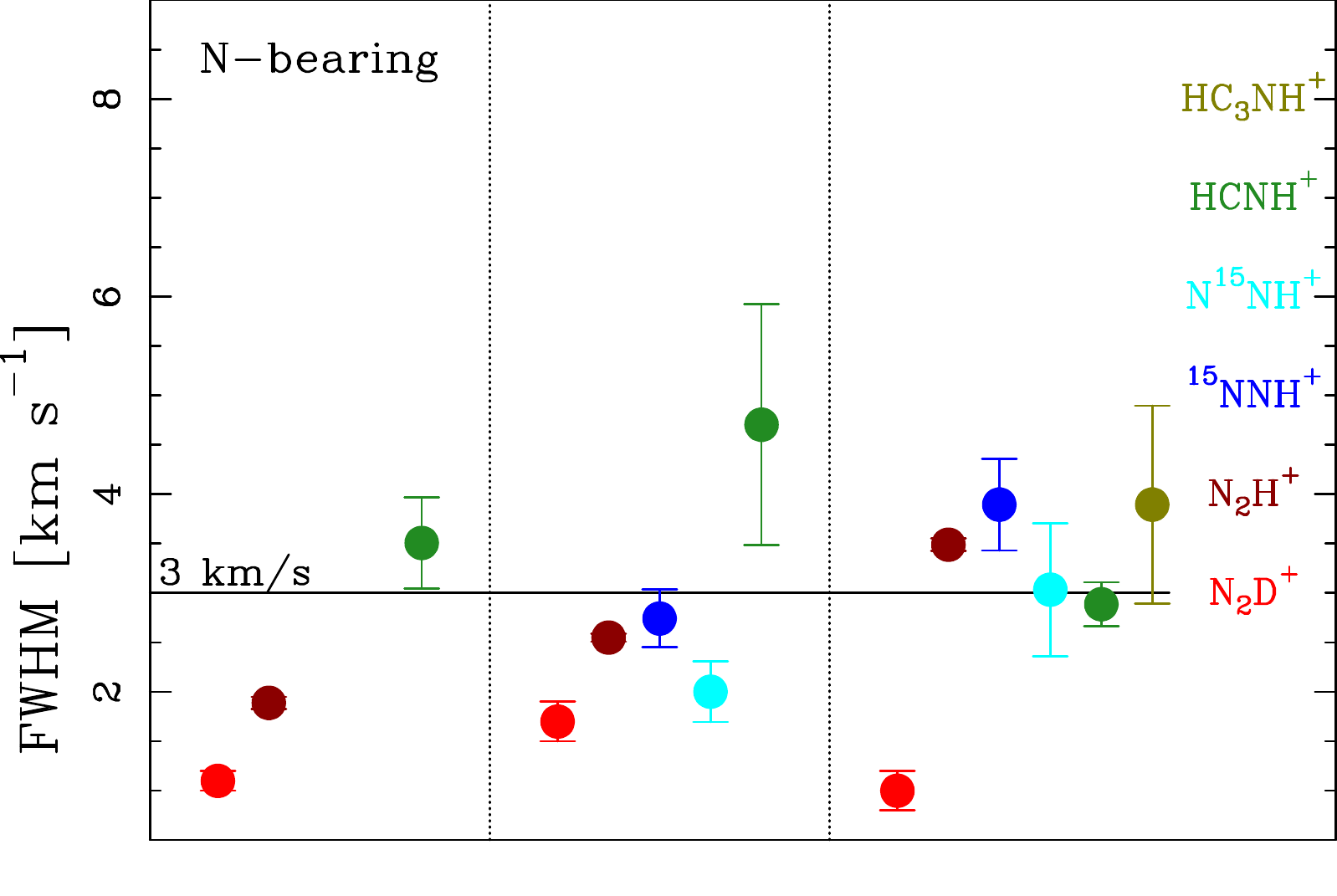}
   \includegraphics[width=7.8cm]{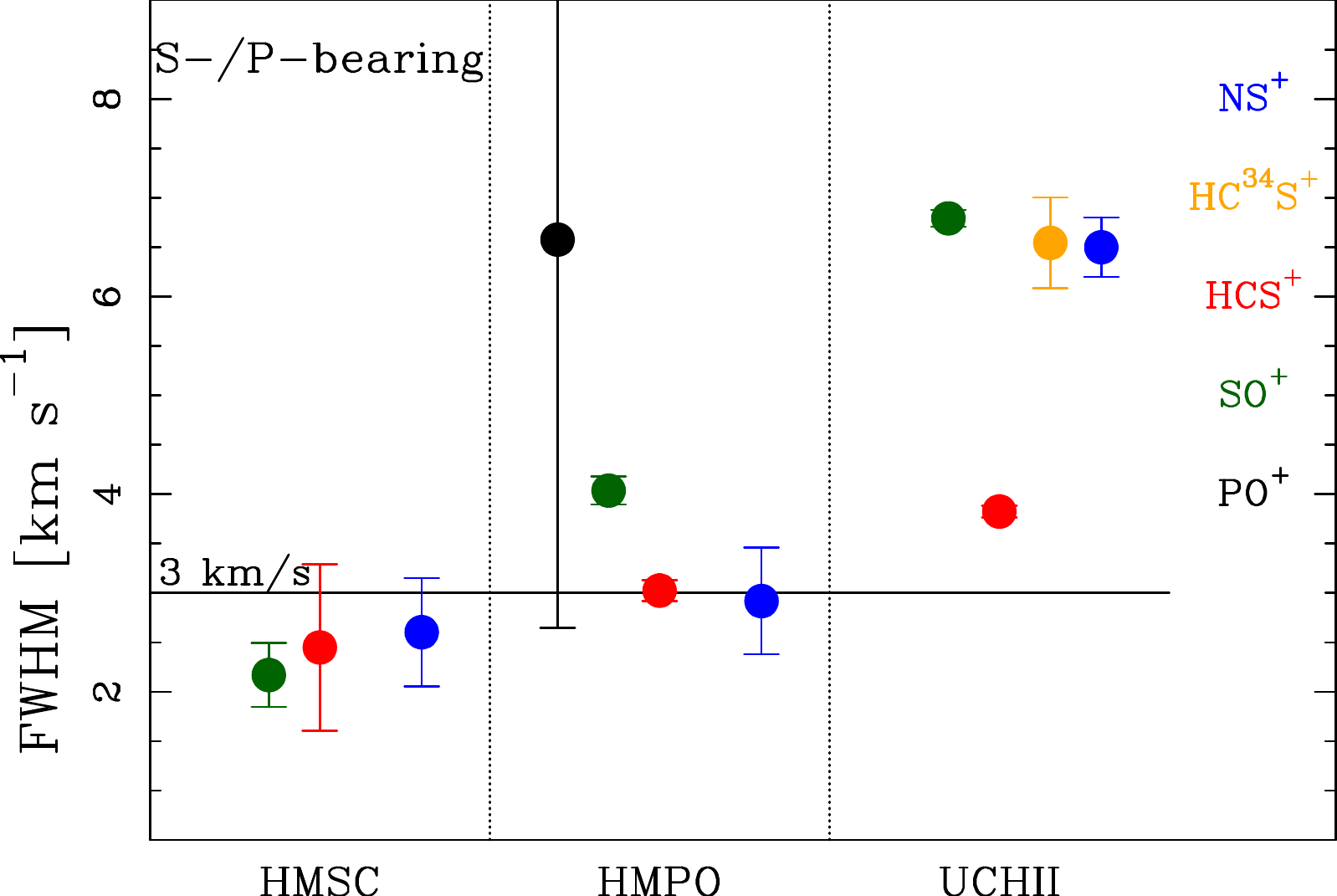}
      \caption{FWHM of, from top to bottom, O-, N-, and S-/P-bearing molecular ions. 
              Points indicate the measured values colour-coded according to the labels to the right.
        The horizontal line indicates a FWHM of 3 \kms.
        For G034, we show only the velocity component that peaks at $\sim 41.5$~\kms.
        For G589, only the narrower velocity component is shown.}
    \label{fig:FWHM}
\end{figure}

\begin{figure}
   \centering
   \includegraphics[width=7.8cm]{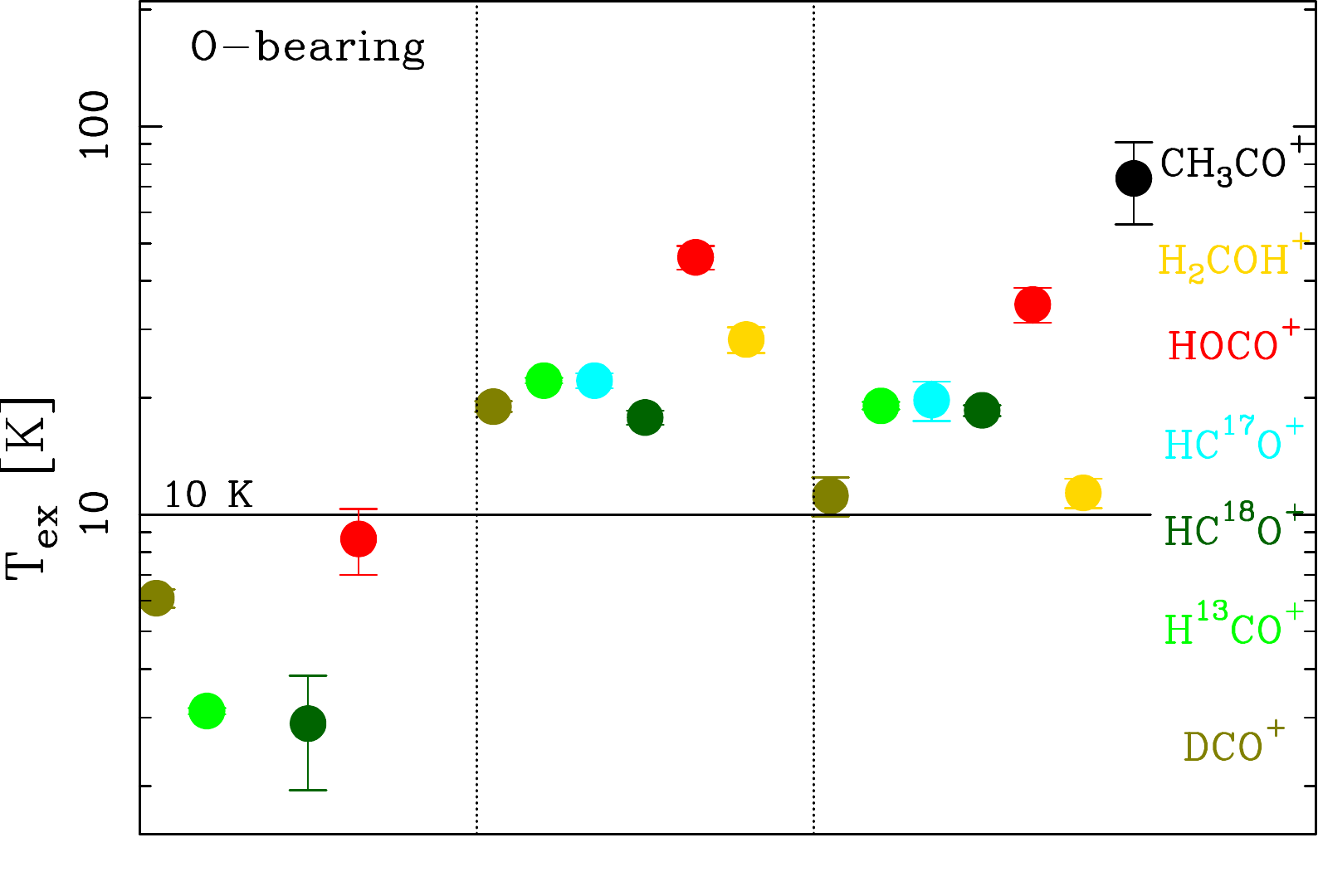}
   \includegraphics[width=7.8cm]{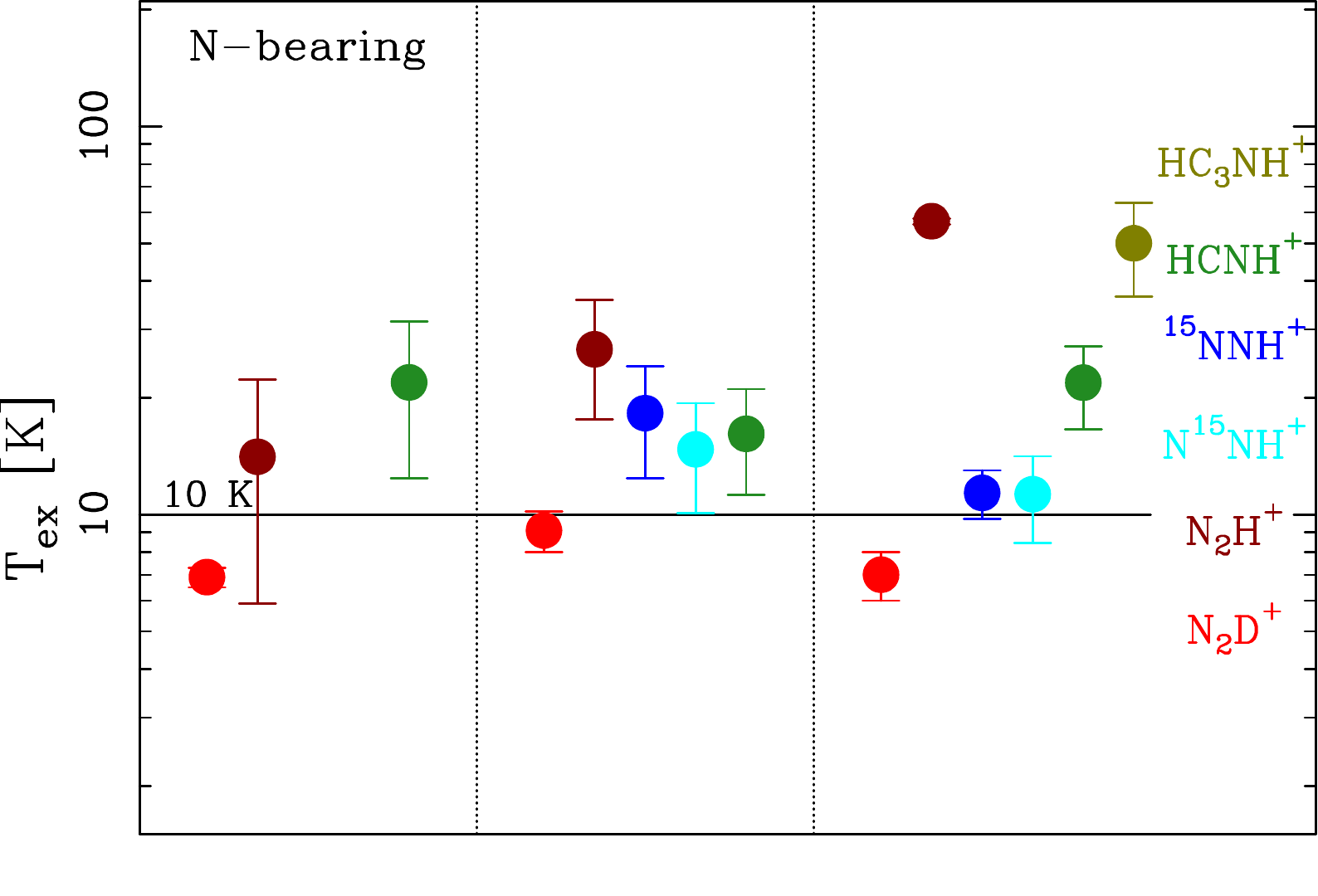}
   \includegraphics[width=7.8cm]{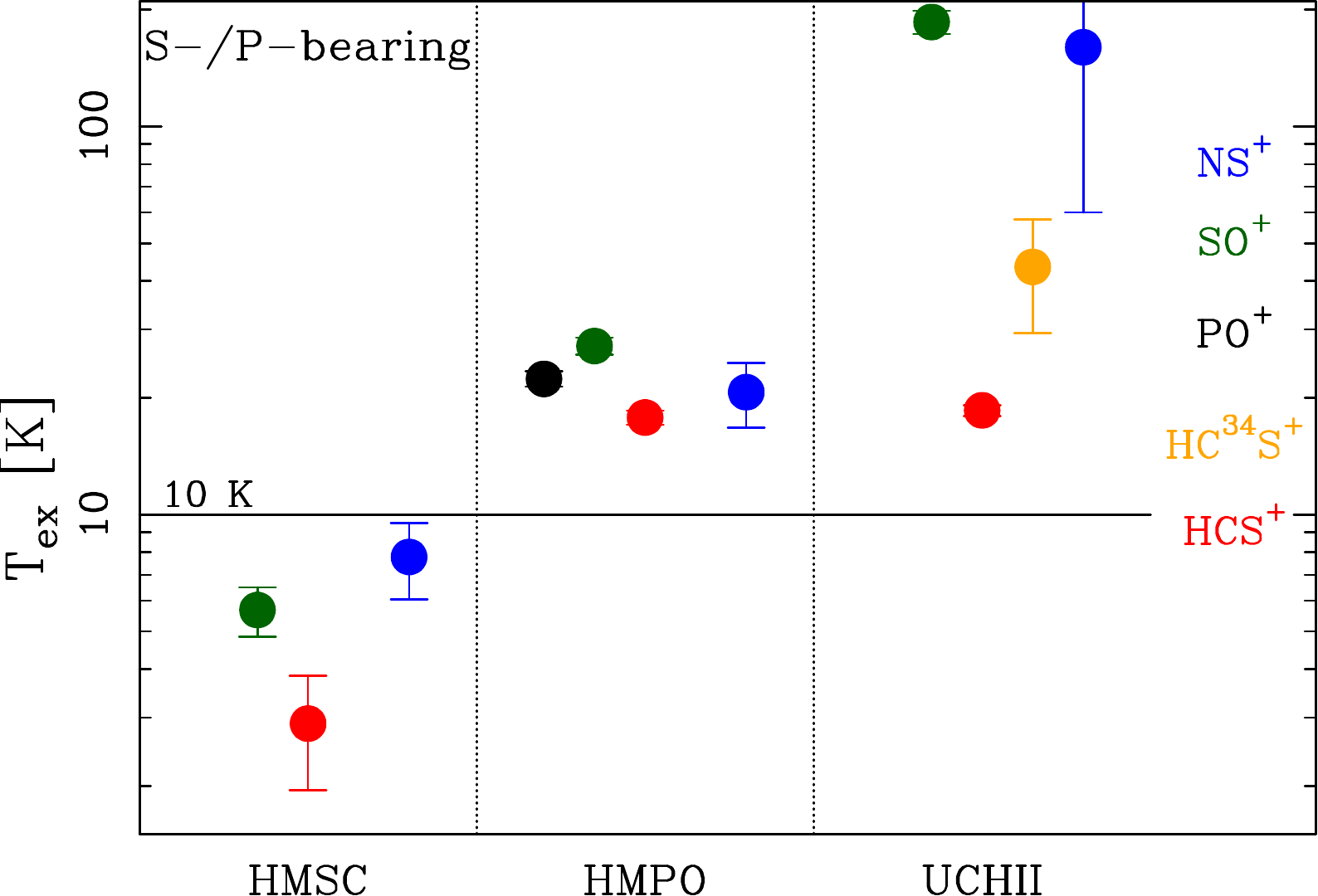}
      \caption{Excitation temperatures of, from top to bottom, O-, N-, and S-bearing molecular ions. 
        Points indicate the measured values colour-coded according to the labels to the right.
    The horizontal line indicates a \Tex\ of 10 K.
    For G034 and G589, only one velocity component is shown as in Fig.~\ref{fig:FWHM}.}
    \label{fig:temperature}
\end{figure}

Table~\ref{tab:physical} shows the best-fit FWHM, $V_{\rm p}$, $N$, and \Tex\ of the detected species. 
An upper limit to the column density of the non-detections computed with {\sc madcuba} is also reported.
As said in Sect.~\ref{sampleandobs}, towards G034 the spectra of most species are best fitted with two velocity components with similar FWHM but different $V_{\rm p}$, while towards G589 H$^{13}$CO$^+$, HC$^{18}$O$^+$, and HCS$^+$ are best fitted with two components with similar $V_{\rm p}$ and different FWHM.
We do not show the fit results for HCO$^+$, which is clearly detected towards all targets (Appendix~\ref{app:A}) but the lines show non-Gaussian profiles that seem to be distorted by large line opacities and possible self-absorption, and are not consistent with the fits (in particular velocities and FWHM) of H$^{13}$CO$^+$, which do exhibit nice Gaussian profiles.

As expected, and as already obtained from the carbon-chain species in paper I, there is an overall enhancement of \Tex\ and FWHM with evolution, although the incremental amount varies significantly between species.
In Figs.~\ref{fig:FWHM} and~\ref{fig:temperature} we show the best-fit FWHM and \Tex, respectively, in each source obtained from all molecular ions studied in this work.
The ions are grouped in O-bearing, N-bearing, and S-/P-bearing species.
For the HMSC, we show only the FWHM and \Tex\ associated with the velocity component at $\sim 41.5$~\kms, as the other component is associated with very similar FWHM and \Tex\ in all tracers (Table~\ref{tab:physical}).
For the UCHII, we only show the narrower velocity component, associated with the bulk material of the cores.

Let us inspect first Fig.~\ref{fig:FWHM}. 
Towards the HMSC, the lines of O-, S-, and P-bearing ions have FWHM almost always narrower than $\sim 3$~\kms, while they are broader than that towards the HMPO and UCHII, reaching values higher than 6~\kms\ in several species.
Instead, the N-bearing ions show FWHMs $\leq 4$~\kms\ even in the evolved stages, and no clear systematic differences among the three targets emerge.
The molecule with the narrowest FWHM in all sources is \D\, and hence it is always associated with the most quiescent gas.
Considering individual species, those for which the FWHM continues to grow significantly from one evolutionary stage to the next one are HOCO$^+$ and the HCO$^+$ isotopologues among O-bearing species, N$_2$H$^+$ and its $^{15}$N isotopologues among N-bearing ones, and SO$^+$ among S-bearing ones.
HOCO$^+$ and SO$^+$ show the clearest increase, suggesting that these ions are the most sensitive to turbulence enhancements.
Inspection of Fig.~\ref{fig:temperature} indicates that S-bearing species show an increase in \Tex\ from $\leq 10$~K in the HMSC stage to $\sim 20-100$~K in the HMPO and UCHII stages (Fig.~\ref{fig:temperature}), O-bearing species show an enhancement in \Tex\ only from the HMSC stage to the HMPO one, while N-bearing ions have \Tex\ almost insensitive to evolution.
Considering again individual species, SO$^+$ and NS$^+$ are the only ones associated with a significant evolutionary increase.

In summary, both Figs.~\ref{fig:FWHM} and ~\ref{fig:temperature} suggest that N-bearing ions preferentially trace gas colder, more quiescent, and less affected by evolution than that traced by S-, P-, and O-bearing ones.

\footnotesize
\begin{center}
\begin{table*}
\small
    \caption{Best-fit line and physical parameters.}
    \label{tab:physical}
            \setlength{\tabcolsep}{2.2pt}
    \begin{tabular}{lcccccccccccccc}
            \hline
            \noalign{\smallskip}
        species  & \multicolumn{4}{c}{G034 (HMSC)\tablefootmark{a}} & & \multicolumn{4}{c}{AFGL (HMPO)} & \multicolumn{4}{c}{G589 (UCHII)\tablefootmark{b}} \\
        \noalign{\smallskip}
        \cline{2-5} \cline{7-10} \cline{12-15}
        \noalign{\smallskip}
        & FWHM & $V_{\rm LSR}$ & Log$_{10}N$ & \Tex\ & & FWHM & $V_{\rm LSR}$ & Log$_{10}N$ & \Tex\ & & FWHM & $V_{\rm LSR}$ & Log$_{10}N$ & \Tex\ \\
            &   (\kms) & (\kms) & (\cmq) & (K) & &   (\kms) & (\kms) & (\cmq) & (K) & &   (\kms) & (\kms) & (\cmq) & (K)  \\
            \noalign{\smallskip}
            \hline
            \noalign{\smallskip}
NS$^+$      &  2.6(0.5)	& 41.7(0.2)	& 11.4(10.9)  & 7.8(1.7)  &   &   2.9(0.5) &	--2.9(0.2)	& 11.7(11.0) & 21(4)   & &  6.5(0.3) &   9.3(0.1) &  12.82(12.5) &   160(90) \\
PO$^+$      &  -- & --	& $\leq 10.6$   &  7.5   &  &   7(4) &	--1.8(1.6) & 11.4(11.1)	& 22.4   &  & -- & --   & $\leq 11.6$ & 36 \\
SO$^+$     &  2.2(0.3)	& 42.0(0.2) & 12.7(12.1)  & 5.7(0.8) &  & 4.0(0.2) &	--3.4(0.1) & 13.26(11.8)	& 27.2(1.4) &  &  6.8(0.1)  & 9.3(0.1)   & 14.7(13.5)	& 186(13) \\
         &  0.9(0.2)	& 43.5(0.1)	& 11.6(11.3)  & 	5.2(1.2) &    &   -- &    -- &  -- &   --  & &  --  &  --  &  --    &    --  \\
HCO$^+$\tablefootmark{c}     &  --	& --	& --  & 	--  &  &  -- & -- & -- & --  & &  --  &  --   &   --    &   --  \\
DCO$^+$     &  1.2(0.1) & 41.7(0.1)	& 11.9(10.9)  & 6.1(0.3)  &   &   2.9(0.1) &	--2.6(0.1) & 12.32(10.7)	& 19.0(0.6) &  &   3.5(0.3)  & 7.7(0.1)   & 11.93(10.9)	& 11(1) \\
         &  0.9(0.2)	& 43.5(0.1)	& 11.6(11.3)  & 	5.2(1.2) &    &   -- &    -- &  -- &   --  & &  --  &  --  &  --    &    --  \\
H$^{13}$CO$^+$   &  1.2(0.2)	& 41.6(0.1)	& 12.8(12.3)  & 3.1(0.1)  &   &   3.2(0.1) & --2.8(0.1) & 12.83(10.9)	& 22.1(0.3) &  &   3.7(0.1)  & 9.0(0.1)   & 13.30(11.6) & 19.1(0.4) \\
         &  1.3(0.2)	& 43.3(0.1) & 11.79(11.3)  & 3.8(0.4)  &  & -- & -- &  --   &  --   & &  12.5(0.2)  & 11.2(0.1)   & 13.34(11.9) & 41(4) \\
HC$^{18}$O$^+$   &  2.5(0.8)	& 42.2(0.2)	& 12.0(12.8)  & 	2.9(0.9)   & &  3.0(0.1) &	--2.9(0.1) & 11.88(10.5)	& 17.8 (0.7) &  &   3.8(0.1)  & 8.8(0.1)   & 12.43(10.8)	& 18.6(0.6) \\
         &  -- &   --   &  --  &   -- &  &   -- &   -- &  --   &  -- &  &  11.2(0.2)  & 11.1(0.13)   & 13.3(13.5) & 470\tablefootmark{d} \\
HC$^{17}$O$^+$   &  --	& --	& $\leq 10.5$   &     3.2 &  & 4.5(0.3) &	--3.1(0.1)	& 11.48(10.3) & 22.1   & &  4.4(0.2)  & 8.7(0.1)   & 11.94(10.6)	& 20(2) \\
HCS$^+$     &  2.8(0.3)	& 42.4(0.1)	& 12.1(11.4)  & 5.4(0.6)  &   &   3.44(0.05) &	--2.3(0.1) & 13.04(11.1)	& 44.8(0.9)  & &   4.2(0.1)  & 9.3(0.1)   & 13.44(11.8) & 60(3) \\
         &  -- &    --    &  --  &   --  & &     -- &     -- &     --   &   -- &  &  13.8(0.4)  & 13.04(0.2)   & 13.2(12.3) & 92(17) \\
HC$^{34}$S$^+$   &  --	& -- &  $\leq 11.2$   &   --     &  & 3 &	--3.34 & 11.7(10.8)	& 50   & &  6.5(0.5)  & 9.0(0.2)   & 12.4(11.5)	& 43(14) \\
N$_2$H$^+$     &  1.9(0.1)	& 41.5(0.1)	& 13.1(12.5)  & 	14(8)   & &   2.6(0.1) &	--2.7(0.1)	& 13.8(13.2)	& 27(9)   & &  3.5(0.1)  & 8.6(0.1)   & 14.50(12.7)	& 57\tablefootmark{e} \\
         &  1.7(0.2)	& 43.3(0.1)	& 12.6(11.9)  & 	6.8(9.1)     &   &   -- &     -- &     --   &   --   & & --  &  --   &   --   &    --  \\
N$_2$D$^+$     &  1.1(0.1) &  41.7(0.1)    & 11.85(10.8)  &  6.9(0.4)   & &   1.7(0.2) & --2.6(0.1) &  11.46(10.6) & 9.1(1.1)   & & 1.0(0.2)  &  7.4(0.1)   &  11.54(10.9)    &    7(1)  \\
           &  1.1(0.2) &  43.5(0.1)    & 11.40(10.8)  &  6.5(0.9)   & &  -- & -- &  -- & --  & & -- &  --  &  --   & --  \\
$^{15}$NNH$^+$   &  --	& --	& $\leq 11.2$   &     --	& &   2.7(0.3) & --3.1(0.1)	& 11.15(10.4)	& 18(6)   & &  3.9(0.5)  & 8.4(0.2)   & 11.42(10.7) & 11(2) \\
N$^{15}$NH$^+$   &  -- & -- & $\leq 10.4$   &  -- & &   2.0(0.3) &	--2.9(0.1)	& 11.1(10.5) & 15(5)   &  & 3.0(0.7)  & 8.2(0.3)   & 11.3(10.8)	& 11(3) \\
HCNH$^+$    &  3.5(0.5)	& 42.1(0.2)	& 13.3(12.4)  & 22(10)  &  &   4.7(1.2) &	--3.3(0.5)	& 13.25(12.7)	& 16(5)  & &   2.9(0.2)  & 8.3(0.1)   & 14.03(12.9)	& 22(5) \\
HOCO$^+$    &  2.9(0.6)	& 41.9(0.2)	& 11.6(11.0)  & 	8.7(1.7)  &   &	 4.6(0.2) &	--2.8(0.1)	& 12.5(11.2)	& 46(3)  & &   7.1(0.7)  & 10.0(0.3)   & 12.6(11.6)	& 35(4) \\
H$_2$COH$^+$   &  --	& --	& $\leq 11.3$   &  --   &  &   7.2(0.5) &	--3.1(0.2)	& 12.95(11.8)	& 28(2)  & &  3.9  & 8.4   & 12.43(11.5)	& 11.4 \\
CH$_3$CO$^+$   &  -- &    --    &  $\leq 11.1$ & --  &  &   -- &	-- & $\leq 12.5$	& --   & & 2.6 & 9.0   & 12.3(11.5)  & 73(17) \\
HC$_3$NH$^+$   &  -- & -- & $\leq 12.0$   & -- & &  -- & -- & $\leq 12.4$ & --  &  & 3.9  & 8.4   & 12.33(11.3)	& 50  \\
            \noalign{\smallskip}
            \hline
            \noalign{\smallskip}
            \end{tabular}
\tablefoot{Values quoted in brackets are uncertainties provided by {\sc madcuba}. The best-fit parameters that do not have an associated uncertainty are fixed. 
\tablefoottext{a}{The spectra of NS$^+$, HC$^{18}$O$^+$, HCS$^+$, HCNH$^+$, and HOCO$^+$ are best-fitted with a single velocity components. The others are best-fitted with two velocity components.}
\tablefoottext{b}{The spectra of H$^{13}$CO$^+$, HC$^{18}$O$^+$, and HCS$^+$ are best-fitted with two (narrow and broad) velocity components. The others are best-fitted with a single velocity component.}
\tablefoottext{c}{HCO$^+$ is clearly detected in all targets but not analysed because the line profiles are complex and distorted (see Sect.~\ref{phys-par};}
\tablefoottext{d}{uncertainty larger than the best-fit value;}
\tablefoottext{e}{Fixed to the value derived in \citet{fontani15a}.}
}
\end{table*}
\end{center}   
\normalsize

\subsection{Molecular fractional abundances}
\label{abundances}

\begin{table}[]
    \centering
        \caption{Fractional abundances of molecular ions.}
    \label{tab:abundances}
    \begin{tabular}{lccc}
 \noalign{\smallskip}
\hline
\noalign{\smallskip}
Species & G034[HMSC] & AFGL[HMPO] & G589[UCHII] \\
\noalign{\smallskip}
\hline
\noalign{\smallskip}
\multicolumn{4}{c}{O-bearing} \\
\noalign{\smallskip}
DCO$^+$    & 3.0$\pm$1.6(-11) &  2.1$\pm$0.8(-11)  & 1.5$\pm$0.7(-12) \\
H$^{13}$CO$^+$  & 1.5$\pm$1.2(-10) &  0.7$\pm$0.2(-10)  & 0.8$\pm$0.3(-10) \\
HC$^{18}$O$^+$  & 2.4$\pm$1.5(-11) &  0.8$\pm$0.3(-11)  & 4.3$\pm$2.1(-11) \\
HC$^{17}$O$^+$  & $\leq 0.82$(-12)  &  3.0$\pm$1.2(-12)  & 1.6$\pm$0.6(-12) \\
HOCO$^+$      & 1.1$\pm$0.6(-11) &  2.9$\pm$1.1(-11)  & 0.8$\pm$0.3(-11) \\
H$_2$COH$^+$  & $\leq 0.54$(-11)  &  0.9$\pm$0.4(-10)  & 4.9$\pm$2.3(-12) \\
CH$_3$CO$^+$  & $\leq 3.2$(-12)   &  $\leq 3.0$(-11)    & 3.8$\pm$1.9(-12) \\
 \noalign{\smallskip}
\hline
\noalign{\smallskip}
\multicolumn{4}{c}{N-bearing} \\
\noalign{\smallskip}
N$_2$H$^+$   & 4.2$\pm$0.3(-10) & 0.6$\pm$0.4(-09) &  0.6$\pm$0.2(-09) \\
N$_2$D$^+$   & 2.3$\pm$1.1(-11) & 2.9$\pm$1.4(-12) &  0.6$\pm$0.4(-12) \\
$^{15}$NNH$^+$ & $\leq 3.7$(-12)   & 1.4$\pm$0.7(-12) &  4.8$\pm$2.7(-13) \\
N$^{15}$NH$^+$ & $\leq 0.7$(-12)  & 1.3$\pm$0.8(-12) &  3.4$\pm$2.4(-13) \\
HCNH$^+$  & 4.9$\pm$2.3(-10) & 1.8$\pm$1.1(-10) &  1.9$\pm$0.8(-10) \\
HC$_3$NH$^+$ & $\leq 2.5$(-11)   & $\leq 2.7$(-11)   &  3.9$\pm$1.7(-12) \\
 \noalign{\smallskip}
\hline
\noalign{\smallskip}
\multicolumn{4}{c}{S-/P-bearing} \\
\noalign{\smallskip}
NS$^+$    & 0.6$\pm$0.4(-11)  & 0.6$\pm$0.3(-11)  &  1.2$\pm$1.0(-11) \\
PO$^+$    & $\leq 0.9$(-12)   & 2.3$\pm$1.9(-12) &  $\leq 0.8$(-12)  \\
SO$^+$    & 1.2$\pm$0.7(-10)  & 1.8$\pm$0.7(-10)  &  0.9$\pm$0.4(-09) \\
HCS$^+$   & 2.8$\pm$1.5(-11)  & 1.1$\pm$0.4(-10)  &  0.8$\pm$0.3(-10) \\
HC$^{34}$S$^+$ & $\leq 3.7$(-12)    & 4.7$\pm$2.2(-12)  &  4.5$\pm$2.2(-12) \\
   \noalign{\smallskip}
\hline
\noalign{\smallskip}
    \end{tabular}
    \tablefoot{Values are in units of $10^x$, with $x$ indicated in brackets.
    Species are grouped in O-, N- and S-/P-bearing molecules as in Fig.~\ref{fig:abb-tot}.}
\end{table}

We derived the molecular fractional abundances with respect to H$_2$, $x$=$N$/\NH, by dividing the molecular column densities
in Table~\ref{tab:physical} by the H$_2$ column densities in Table~\ref{tab:sources}. 
The results are listed in Table~\ref{tab:abundances}.
Because \NH\ is an average value over an angular region of 28\asec, and $N$ is computed assuming the emission fills the beam of the CHEMICO observations, $x$ is likewise an average value over 28\asec. 
In spectra where two velocity components are detected within the CHEMICO beam  (see Sect.~\ref{sampleandobs} and Table~\ref{tab:physical}), we summed their column densities to compute $x$. 
Figure~\ref{fig:abb-tot} shows the abundances measured in the three CHEMICO sources for O-bearing, N-bearing, and S-/P-bearing ions separately.
Most species do not show obvious trends with evolution.
Because the H$_2$ column density increases from the HMSC to the UCHII stage (Table~\ref{tab:sources}), this result indicates that the molecular column densities also increase, thus keeping the majority of the abundance trends as flat.
In the following subsections, we discuss the few species associated with an evolutionary trend in more detail.

\subsubsection{Species with a clear abundance trend: SO$^+$, \D, and DCO$^+$}
\label{n2dpanddcop}

We propose as species associated with a clear abundance trend those for which the measured abundance continuously increase, or decrease, from one stage to the following one, and the change is significant (considering uncertainties).
These are SO$^+$, \D, and DCO$^+$.
The increasing trend of $x$(SO$^+$), already noticed by \citet{fontani23}, is consistent with an origin in dissociative protostellar shocks, being the most efficient reaction ${\rm S^+ + OH \rightarrow SO^+ + H}$ \citep[e.g.,][]{ned89,turner92,rivilla22}.
\citet{turner94} also highlights the need for S$^+$ to produce SO$^+$ in a reaction with shock-enhanced OH.
Another channel proposed to efficiently form SO$^+$ is ${\rm S^+ + O_2 \rightarrow SO^+ + O}$
\citep{stauber05}, which still requires S$^+$ produced by shocks or far-UV radiation.
Comparing the behaviour of SO$^+$ with that of the other S-bearing ions, HCS$^+$ is the protonated form of CS that does not require shock conditions to be formed, while NS$^+$, detected first in the interstellar medium (ISM) by \citet{cernicharo18}, can also be formed by S$^+$ through ${\rm S^+ + NH \rightarrow NS^+ + H}$ \citep[e.g.,][]{riviere-marichalar19}.
However, in cold conditions, ion-neutral formation reactions between NS and atomic cations \citep{lec19}, or between atomic N and S-bearing cations \citep[e.g.,][]{cernicharo18}, are also possible.
Therefore, the fact that SO$^+$ is originated solely from S$^+$ could explain why this is the only S-bearing species that shows a clear evolutionary enhancement.

The decrease with evolution of $x$(\D) and $x$(DCO$^+$) is obviously due to the temperature enhancement with time (see paper I). 
In fact, both \D\ and DCO$^+$ are formed efficiently by reactions with H$_2$D$^+$ (namely ${\rm N_2+H_2D^+\rightarrow N_2D^++H_2}$ and ${\rm CO+H_2D^+\rightarrow DCO^++H_2}$, respectively), whose abundance is boosted in the early cold ($T\leq 20$~K) phases by the highly exothermic reaction \citep{watson76}: 
\begin{equation}
{\rm H_3^++HD\rightarrow H_2D^+ + H_2} + \Delta E\;,
\label{eq:h2dp}
\end{equation}
if H$_2$ is mainly in the para- form.
If there is a small amount of ortho-H$_2$, then H$_2$D$^+$ is destroyed by it and its abundance is lowered, followed by those of both DCO$^+$ and \nndp\ \citep[e.g.][]{flower06}.
Once the temperature raises above $\sim 20$~K, ${\rm H_2D^+}$ is rapidly destroyed by the reverse reaction (\ref{eq:h2dp}), thus lowering the abundance of its daughter species \D\ and DCO$^+$.
For \D, our finding is perfectly in line with this scenario, already widely demonstrated by previous observations towards both low- and high-mass star-forming regions in different evolutionary stages \citep[e.g.,][]{emprechtinger09,fontani11,fontani25a}.
DCO$^+$ is more elusive because its abundance decreases with evolution in a less pronounced fashion with respect to \D, as indicated by the marginal decrease from the HMSC to the HMPO stages.
This finding also agrees with previous studies that indicate that the DCO$^+$ abundance can remain high also in evolved regions \citep[e.g.,][]{roueff07,li17} owing to efficient production in warm gas by:
\begin{equation}
    {\rm CH_3^+ + HD \rightarrow CH_2D^+ + H_2} + \Delta E\;,
    \label{eq:ch2dp}
\end{equation}
followed by
\begin{equation}
   {\rm CH_2D^+ + O \rightarrow DCO^+ + H_2}\;.
   \label{eq:dcop}
\end{equation}
As discussed in \citet{roueff07} (see also Favre et al.~\citeyear{favre15}), {\rm $\Delta E$} in reaction~(\ref{eq:ch2dp}) is $\sim 650$~K \citep{roueff13}.
Therefore, CH$_2$D$^+$ can become/remain enhanced even at temperature $\gg 100$~K, and the DCO$^+$ formation channel through reaction~(\ref{eq:dcop}) can compensate for destruction due to reverse reaction~(\ref{eq:h2dp}).


Linked to what discussed above, the D/H ratio in \nndp\ and DCO$^+$ decrease with evolution.
From the abundances in Table~\ref{tab:abundances}, the D/H ratio derived from the \nndp/\nnhp\ abundance ratio is $\sim 0.06$, $\sim 0.005$, and $\sim 0.001$ in G034, AFGL, and G589, respectively. 
This finding is in agreement with the decrease of an order of magnitude from the HMSC phase to the later phases found in \citet{fontani11}.
However, in that study the \nndp\ column density towards G034 is higher by an order of magnitude.
To understand the discrepancy, we have tried to reproduce the column density of \nndp\ in \citet{fontani11} using their approach applied to the CHEMICO data.
To do this, we used only the $(2-1)$ transition (which is the only one observed in Fontani et al.~\citeyear{fontani11}), and we fixed both \Tex\ and the source size to the values assumes in \citet{fontani11}, namely \Tex=16 K and 6.5\asec, respectively.
With these assumptions we indeed derive a total column density of $\sim 10^{13}$~\cmq, consistent with the one in \citet{fontani11}.
However, first our multiline analysis provides \Tex$\sim 6-7$~K for \nndp, smaller than that assumed in \citet{fontani11}.
Moreover, the high-angular resolution data of \citet{tan13} showed that the angular emission size of \nndp\ is $\sim 15$\asec, larger than the size assumed in \citet{fontani11}.
Therefore, our method, in which we assume that the emission fills the telescope beam and \Tex\ is derived from a multiline analysis, is more accurate.

The D/H ratio derived from DCO$^+$ is 0.003, 0.005, and 0.0003, for G034, AFGL, and G589, respectively.
This indicates a significant decrease only from the HMPO to the UCHII stage.
Therefore, the difference in the D/H ratio from \nnhp\ and HCO$^+$ can discriminate between starless and protostellar stage.
Finally, the $^{14}$N/$^{15}$N ratio derived from \nnhp\ towards AFGL is $\sim 450$ and towards G589 is $\sim 700$, consistent within the uncertainties with previous measurements \citep{fontani15b}.
The different trends found for the D/H and $^{14}$N/$^{15}$N ratios in the same molecule indicate once more that the two isotopic ratios do not evolve in the same way \citep[e.g.,][]{colzi18}.

\subsubsection{Species associated with an abundance peak in the HMPO stage: PO$^+$, H$_2$COH$^+$, and HOCO$^+$}

\begin{figure}
   \centering
   \includegraphics[width=7.8cm]{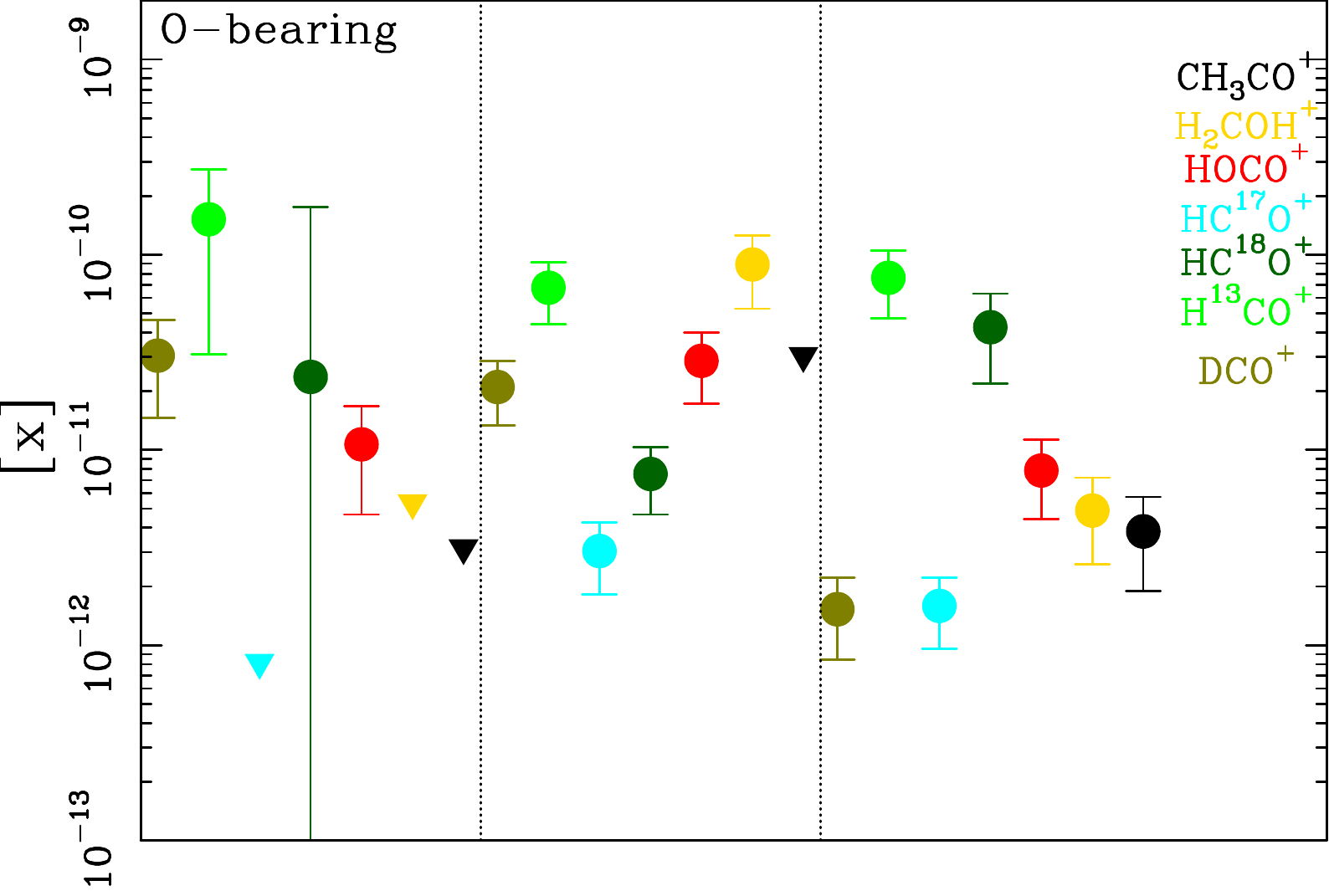}
   \includegraphics[width=7.8cm]{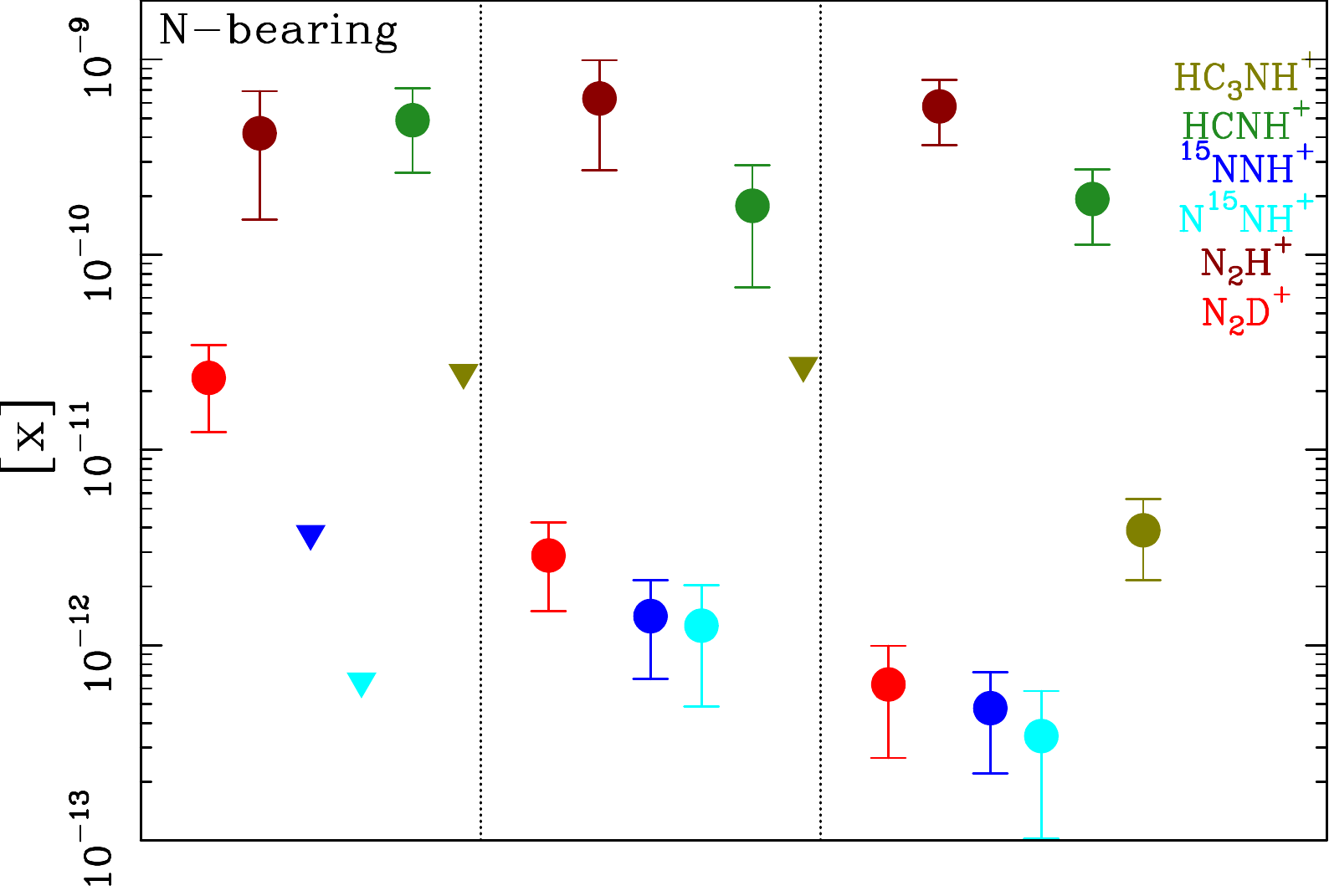}
   \includegraphics[width=7.8cm]{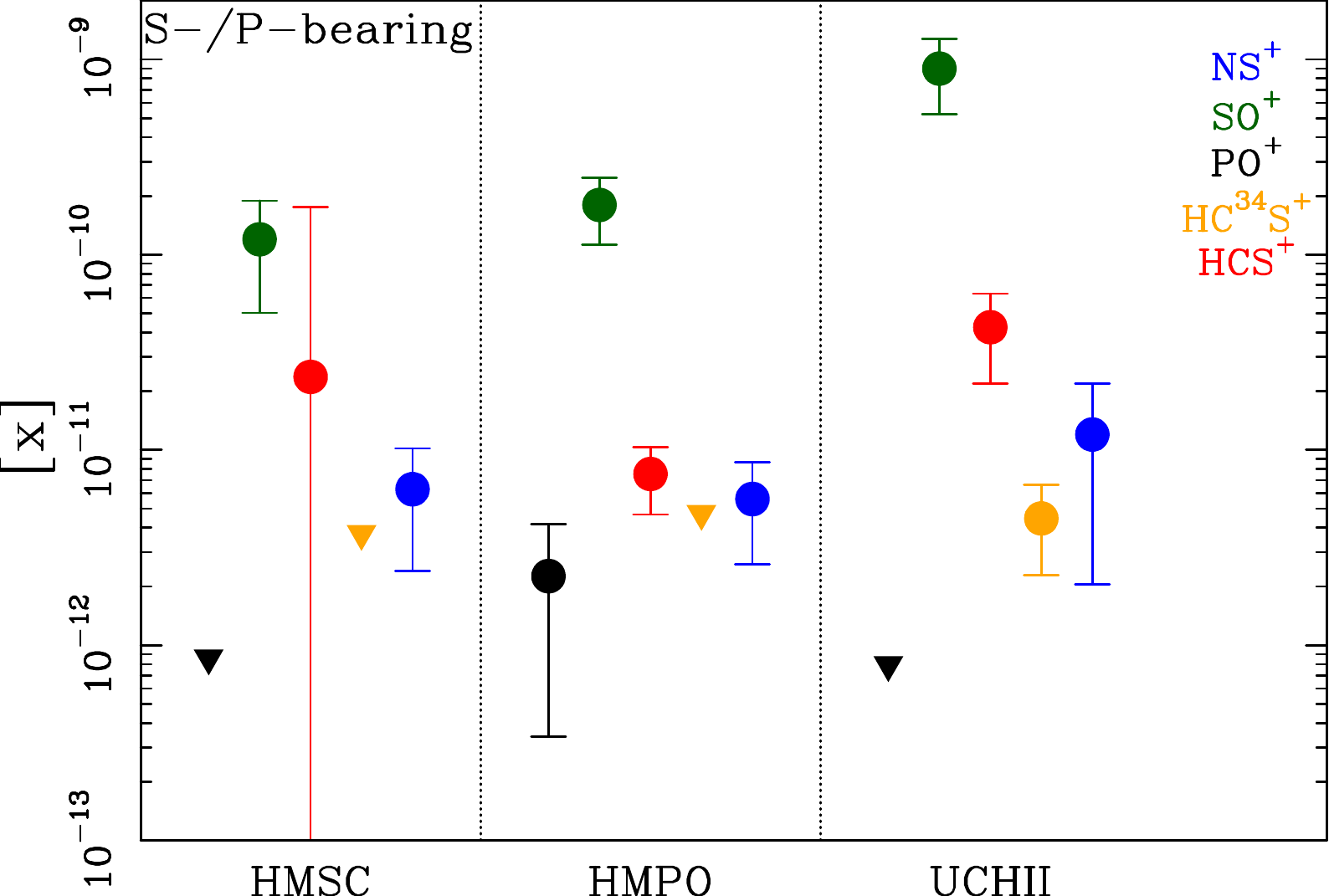}
      \caption{Fractional abundances of, from top to bottom, O-, N-, and S-/P-bearing molecular ions. 
              Filled circles indicate the measured values colour-coded according to the labels in the top right corner.
              Triangles correspond to upper limits.}
    \label{fig:abb-tot}
\end{figure}

The abundances of three ions reach the maximum value at the HMPO stage: PO$^+$ (considering upper limits), H$_2$COH$^+$, and, HOCO$^+$.
The trend for HOCO$^+$ is tentative as its abundance in the HMPO stage is only a factor of a few higher than those in the HMSC and UCHII stages.
The PO$^+$ abundance greatly increases in shocked regions with high cosmic-ray ionisation rates \citep{rivilla22,scibelli25}.
Therefore, our result is consistent with a scenario in which PO$^+$ is originated by CRs locally accelerated by protostellar activity \citep{padovani16,geo18}.
Both H$_2$COH$^+$ and HOCO$^+$ are protonated species predominantly formed by proton transfer from H$_3^+$ or other proton donors of the type XH$^+$, and destroyed by electrons upon dissociative recombination (see e.g. the discussion in Bacmann et al.~\citeyear{bacmann16} for H$_2$COH$^+$ and Fontani et al.~\citeyear{fontani18} for HOCO$^+$).
Thus, their formation could be enhanced at the HMPO stage owing to the higher ionisation of the gas, like PO$^+$.
However, the abundances of other protonated species predominantly formed by proton transfer, such as HCS$^+$, N$_2$H$^+$, HCO$^+$, and HCNH$^+$, remain constant within the uncertainties in the three phases (Fig.~\ref{fig:abb-tot}).
In addition to proton transfer, HOCO$^+$ could be produced in the HMPO stage also from gaseous OH, released in the gas phase upon grain mantle evaporation, via reaction ${\rm HCO^+ + OH \rightarrow HOCO^+ + H}$ \citep[e.g.,][]{vastel16}.
Similarly, a non-negligible source of H$_2$COH$^+$ in the HMPO stage can be dissociation of \METH, another species abundantly released in the gas from grain mantles, through UV photons \citep[e.g.,][]{pilling07}.
These processes could explain the enhancement of both species from the HMSC to the HMPO stage, but not the subsequent decrease in the UCHII stage.

\section{Discussion}
\label{discu}

\subsection{Protonated-to-neutral abundance ratios}
\label{affinity}

\begin{figure}
   \centering
   \includegraphics[width=9cm]{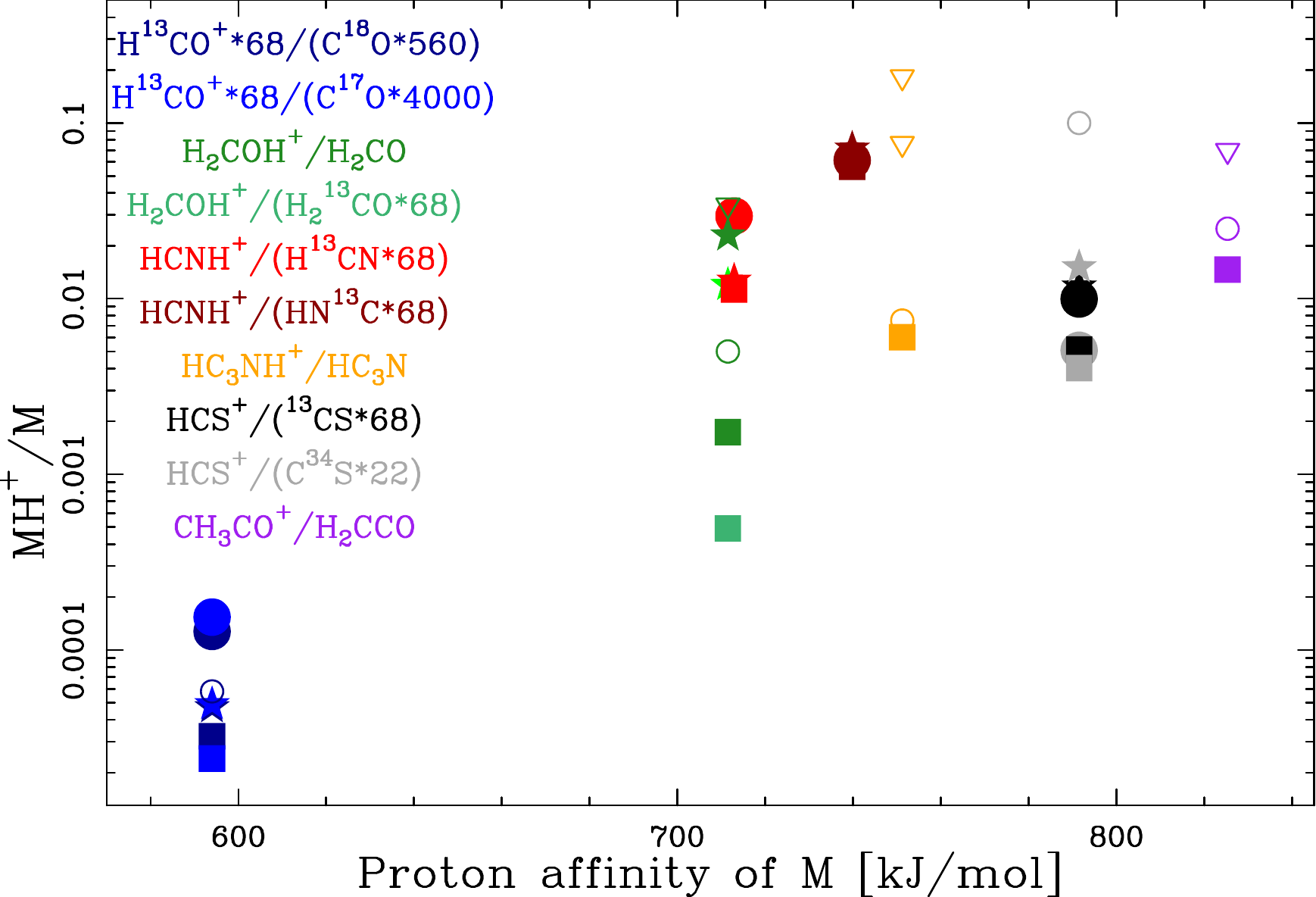}
      \caption{Column density ratio MH$^+$/M as a function of the M proton affinity.
      Proton affinities are taken from the National Institute of Standards and Technology (NIST) chemistry WebBook (\url{https://webbook.nist.gov}).
      The colours indicate the different species according to the description in the top left corner.
      Ratios measured for G034 (HMSC) are indicated by circles, for AFGL (HMPO) by stars, and for G589 (UCHII) by squares.
      Open triangles indicate upper limits.
      The open circles indicate the ratios reported by \citet{agundez22} observed towards the low-mass dense cold cloud TMC--1.}
    \label{fig:affinity}
\end{figure}

Several molecular ions analysed in this work are protonated forms of abundant neutral species, such as CO, CS, HCN, HNC, H$_2$CO, H$_2$CCO, and HC$_3$N.
As discussed in \citet{agundez22}, if M is a neutral species, and if proton transfer to M is the main formation pathway to produce MH$^+$, the species with the higher proton affinity should show MH$^+$/M column density or abundance ratios higher than those with lower proton affinity.
To test this, in Fig.~\ref{fig:affinity} we show the MH$^+$/M ratios of the neutral species mentioned above as a function of the proton affinity of M.
The ratios were computed dividing the column densities of the protonated species in Table~\ref{tab:physical} by those of the corresponding neutral species in Table~\ref{tab:neutral} of Appendix~\ref{appNeutral}, identified and analysed in the CHEMICO dataset following the same method described in Sect.~\ref{sampleandobs}.
The listed column densities are the sum of the multiple velocity features eventually present in the spectra (Table~\ref{tab:physical}).
For species associated with optically thick lines, namely CO, HCO$^+$, CS, HCN and HNC, we converted the column densities of the less abundant isotopologues to the main one by applying the following isotopic ratios: $^{12}$C/$^{13}$C=68~\citep{wilson99}, $^{16}$O/$^{18}$O=560~\citep{wilson99}, $^{16}$O/$^{17}$O=2000~\citep{wilson99}, $^{32}$S/$^{34}$S=22~\citep{lodders03}.

The trend shown in Fig.~\ref{fig:affinity} is consistent with the expected increase of the MH$^+$/M ratio with proton affinity.
The same finding was first noticed by \citet{agundez22} in a larger number of molecular ions towards the cold dense cores TMC--1 and L483, hence only towards cold dark cloud conditions.
The main novelty of our study is that the increasing trend applies also to warm protostellar conditions.
Overall, Fig.~\ref{fig:affinity} indicates mainly that neutral species with proton affinity larger than that of CO are associated with MH$^+$/M abundance ratios in between $10^{-2}$ and 0.1 (except H$_2$COH$^+$/H$_2$CO), while HCO$^+$/CO is of the order of $10^{-4}$ in all targets.
This can be explained by the fact that HCO$^+$
can be a proton donor for all species with higher affinity, causing an increase in the abundance of their protonated counterparts at the cost of that of HCO$^+$.

\subsection{Evolution of the ionisation fraction}
\label{ionisation}

The ionisation fraction $x(e)$, defined in molecular gas as the electron fractional abundance with respect to H$_2$, $x(e) = n(e^-)/n({\rm H_2})$,
is a fundamental parameter for both the physical and chemical evolution of star-forming regions.
It regulates coupling between gas and magnetic field, and thus, during collapse, the ambipolar diffusion process \citep[e.g.,][]{shu87}, which in turn influences the efficiency and timescale of star-formation.
Moreover, it regulates the ion-neutral reaction rates, which are critical in cold cloud conditions.
In dense star-forming regions, the ionisation fraction is commonly estimated indirectly via abundances of molecular ions combined with predictions of chemical models \citep[e.g.,][]{langer85}. 
A popular simplified approach in cold conditions (\Tk $\leq 20$~K) and at intermediate H$_2$ volume densities ($n({\rm H_2})\sim 10^4-10^5$~\cmc) links $x(e)$ at steady-state to the molecular abundance ratio DCO$^+$/HCO$^+$=$R_{\rm D}$ \citep{caselli98,miettinen11,pineda24}.
In particular, assuming a kinetic temperature of 10~K, \citet{caselli98} give this simple expression for the electron fraction:
\begin{equation}
x_{\rm C}(e)=\frac{2.7\times 10^{-8}}{R_{\rm D}}-\frac{1.2\times 10^{-6}}{f_{\rm D}}\;,
    \label{eq:ionisation-cas}
\end{equation}
where $f_{\rm D}$ is the ratio between canonical and observed gas-phase abundance of oxygen, which quantifies the fractional amount of O in the form of ices.
$f_{\rm D}$ of O can be assumed to be equal to that of CO, the most abundant O-bearing molecule, which can be estimated as the ratio between the canonical CO abundance with respect to H$_2$ and the observed one.
The reference CO abundance in the local ISM shows a measured spread between $\sim 0.4\times 10^{-4}$ and $2.7\times 10^{-4}$ \citep{punanova22}.
We assumed $2.7\times 10^{-4}$ \citep{lacy94} obtained towards the Orion molecular cloud NGC 2024 because the other estimates were derived towards low-mass star-forming regions Taurus and Ophiuchus, thus less similar to our high-mass targets.
The measured CO column densities were derived analysing either \CII\ or \CIII\ lines (Table~\ref{tab:neutral}) considering isotopic ratios of $^{16}$O/$^{18}$O=560 and $^{16}$O/$^{17}$O=1000, respectively \citep{wilson99}.
The \CII\ and \CIII\ column densities were derived from the CHEMICO dataset following the same methodology adopted for ions (Sect.~\ref{sampleandobs}).
Measured $f_{\rm D}$, obtained either from C$^{18}$O and C$^{17}$O, are in between $\sim 1$ for G589 (i.e. no CO depletion) and $\sim 3-4$ for G034 (Table~\ref{tab:ionisation-luo}).
The values computed from the two isotopologues are consistent among them.

Because our sources are warmer than 10~K (Table~\ref{tab:sources}), in this work we update the approach in \citet{caselli98} considering the explicit dependence of reaction rates on the temperature.
We also include the ortho-to-para ratio of H$_2$, which also varies with temperature.
The updated expression for the electron fraction is:
\begin{equation}
\begin{split}
x_{\rm F}(e)= & \left(\frac{T}{20}\right)^{0.65}\cdot \Bigl[\frac{5.7\times 10^{-8}}{R_{\rm D}}-1.1\times 10^{-7} \\
& -5.7\times 10^{-3}\cdot \frac{\exp(-232/T)
+ opr \cdot \exp(-61.5/T)}{1+opr} \\
& -\frac{4.2\times 10^{-7} \cdot \exp(-1.4/T)\cdot (T/20)^{-0.16}+4.1\times 10^{-7}}{f_D} \Bigr]\;, \\
\end{split}
\label{eq:ionisation-noi}
\end{equation}
where $T$ is the gas temperature (for which we adopt \Tk\ in Table~\ref{tab:sources}), and $opr$ is the ortho-to-para ratio of H$_2$.
We calculated $opr$ from \citet{faure19} considering the \Tk\ in Table~\ref{tab:sources}.
The values are obtained from the chemical balance of a steady state model including a basic $opr$ chemistry for H$_2$, H$_3^+$, and H$_2$D$^+$ from the reaction rates given in \citet{hugo09}.
As shown in Fig.~1 in \citet{faure19}, the modelled $opr$ value is $10^{-3}$ for $T$ lower than 20k (i.e. for G034). 
For higher $T$, the $opr$ rapidly reaches thermal equilibrium, thus the values are calculated in thermal equilibrium.
The derived values of $x_{\rm F}(e)$ are listed in Table~\ref{tab:ionisation-luo}.
The detailed derivation of Eq.~(\ref{eq:ionisation-noi}) is described in appendix~\ref{app:opr}.
We also checked whether introducing a CO abundance tailored for the Galactocentric distance\footnote{We assumed the CO Galactocentric gradient in \citet{fontani06}, as done in \citet{miettinen11}.} of the three sources introduces significant changes in the results, and found negligible variations of $\sim 1-2\%$.

\begin{table*}[]
    \centering
    \caption{Electron fractions, cosmic-ray ionisation rates, and related parameters.}
     \setlength{\tabcolsep}{2pt}
    \begin{tabular}{lccccccccccccc}
    \hline
    \noalign{\smallskip}
   source   & $opr$\tablefootmark{a} & $R_{\rm D}$\tablefootmark{b} & $f_{\rm D}^{18}$\tablefootmark{c} & $f_{\rm D}^{17}$\tablefootmark{c} & $n$(H$_2$)\tablefootmark{d} & $x_{\rm C}(e)^{18}$ &  $x_{\rm C}(e)^{17}$  &  $x_{\rm F}(e)^{18}$  & $x_{\rm F}(e)^{17}$ & $x_{\rm L}(e)^{18}$    &  $x_{\rm L}(e)^{17}$  &  $\zeta_{\rm L}^{18}$  & $\zeta_{\rm L}^{17}$  \\
       \noalign{\smallskip}
			& ($10^{-3}$) & ($10^{-3}$) &     &     & ($10^{5}$~cm$^{-3}$) & ($10^{-5}$) & ($10^{-5}$) & ($10^{-5}$) & ($10^{-5}$) &  ($10^{-8}$) & ($10^{-8}$) & ($10^{-16}$~s$^{-1}$) & ($10^{-16}$~s$^{-1}$) \\
\noalign{\smallskip}
\hline
\noalign{\smallskip}
G034[HMSC]  &  1     &  2.9       & 3.3 & 4.0 & 0.34     &    0.9   & 0.9      &  1.6     & 1.6      & 1.5      & 1.5      &  0.9      & 0.9      \\
AFGL[HMPO]  &  3.2     &  4.6       & 2.8 & 2.9 & 1.3      &    0.6   & 0.6      &  1.1     & 1.1      & 1.1      & 1.1      &  5.3      & 5.2      \\
G589[UCHII] &  25      &  0.3       & 0.7 & 1.3 & 10       &    8.9    & 9.0      &  22     & 22      & 9.8 & 7.9 & 1100 & 590 \\
\noalign{\smallskip}
\hline
    \end{tabular}
    \tablefoot{$x(e)$ is derived from the three methods described in Sect.~\ref{ionisation}.
    $\zeta$ is derived only following \citet{luo24}.
    Uncertainties are in between 30 and 60$\%$.
    Parameters labelled with superscript $^{18}$ and $^{17}$ are derived from $x{\rm (CO)}$ computed from either \CII\ or C$^{17}$O, respectively.
    \tablefoottext{a}{From \citet{faure19} at the kinetic temperature of the sources (Table~\ref{tab:sources}). Values are in units of $10^x$, with $x$ indicated in brackets.;}
    \tablefoottext{b}{From $N$(DCO$^+$) and $N$(H$^{13}$CO$^+$) listed in Table~\ref{tab:physical}. The latter was converted to $N$(HCO$^+$) by multiplying it for $^{12}$C/$^{13}$C=68. For G034[HMSC], the column densities of the two velocity components were summed;}
    \tablefoottext{c}{Derived from both $N$(C$^{18}$O) ($f_{\rm D}^{18}$) and $N$(C$^{17}$O) ($f_{\rm D}^{17}$) in Table~\ref{tab:neutral}, converted to $N$(CO) by multiplying for 560 and 2000, respectively. For spectra with multiple velocity features, the column densities of the two velocity components were summed.}
    \tablefoottext{d}{derived from $N$(H$_2$) (Table~\ref{tab:sources}) on an angular scale of 28\asec.}
    }
    \label{tab:ionisation-luo}
\end{table*}

Both $x_{\rm C}(e)$ and $x_{\rm F}(e)$, and all paramereters used to derive them, are shown in Table~\ref{tab:ionisation-luo}.
As previously noted, the differences between the estimates obtained from \CII\ or \CIII\ are negligible, while those obtained with the two methods (Eqs.~\ref{eq:ionisation-cas} and \ref{eq:ionisation-noi}) are significant, as $x_{\rm F}(e)$ is systematically higher than $x_{\rm C}(e)$ by factors $\sim 2-2.5$.
In any case, both methods lead to unrealistically high electron fractions in the range $10^{-5}-10^{-4}$.
In appendix~\ref{app:opr} we also derive an expression of the cosmic-ray ionisation rate, $\zeta$, linked to $x_{\rm F}(e)$.
However, computed $\zeta$ are also unrealistically high, and we do not list them in Table~\ref{tab:ionisation-luo}, nor elsewhere in the paper.
These results indicate that the \citet{caselli98} method, even after introducing our updates, contains assumptions inappropriate for the physical conditions of our sources.
For example, the method is most reliable in regions where CO is significantly depleted, while the gas probed by our observations shows small or irrelevant CO depletion ($f_{\rm D}$ in Table~\ref{tab:ionisation-luo}).
A similar recent approach based on the chemistry of \nnhp, \nndp and DCO$^+$ was developed by \citet{socci26}, but also this method is designed to simulate the conditions of dense gas at temperatures below $\sim $20~K, where deuteration is essential, and therefore not appropriate at least for two of our three sources (see also Redaelli et al.~\citeyear{redaelli24}, who discuss the range of validity of the various analytic methods).

Instead, the method proposed by \citet{luo24} is specifically designed for warmer gas conditions.
The method uses simplified steady-state chemistry in which the dominant ion species are HCO$^+$ and N$_2$H$^+$, and both deuteration and CO depletion are much less important.
According to this approach, $x(e)$ is expressed 
as a function of three coefficients, $\beta$, $\gamma$, and $\theta$, as:
\begin{equation}
    x_{\rm L}(e)=\frac{1}{n({\rm H_2})}\frac{\beta + \sqrt{\beta^2+4\gamma\theta}}{2\gamma}\;.
    \label{eq:luo}
\end{equation}
Coefficients $\beta$, $\gamma$, and $\theta$ are given in Eqs.~(8), (9), and (10), respectively, of \citet{luo24}, and contain adopted or measured volume densities of O, CO, N$_2$, \nnhp, and \hcop, as well as the reaction rates listed in their Table B.1, and
taken from the UMIST database \citep[][]{millar24}.
In our work, to be consistent with the derivation of $x_{\rm F}(e)$ (Appendix~\ref{app:opr}), we decided to take the reaction rates that enter coefficients $\beta$, $\gamma$, and $\theta$ from the KInetic Database for Astrochemistry \citep[KIDA][]{wakelam24}.
In general, reaction rates from KIDA and UMIST are consistent among them at the temperatures of our sources, with differences below $\sim 10 \%$.
However, we spotted a significant difference.
The rate of reaction ${\rm N_2H^+ + O \rightarrow OH^+ + N_2}$ ($k_{\rm R10}$ in Luo et al.~\citeyear{luo24}) in UMIST misses the factor ${\rm exp(-3400/T)}$ present in KIDA.
Because the reaction is endothermic, as demonstrated by the sum of enthalpies difference between the products and the reactants (see NIST WebBook), the KIDA expression is the correct one.
By including this factor, 
coefficient $\theta$ in Eq.~(\ref{eq:luo}) approaches zero, and thus Eq.~(\ref{eq:luo}) becomes:
\begin{equation}
\begin{split}
x_{\rm L}(e)=  & \frac{1}{n({\rm H_2})}\frac{\beta}{\gamma} =   & 
             \frac{\frac{n({\rm N_2H^+})}{n({\rm N_2})\cdot k_{\rm R4}}\cdot [n{\rm (O)}\cdot \frac{k_{\rm R5}\cdot k_{\rm R9}}{k_{\rm R13}}+n{\rm (CO)}\cdot k_{\rm R8}]}{n({\rm H_2})\cdot [1-\frac{n({\rm N_2H^+)}\cdot k_{\rm R9}}{n({\rm N_2})\cdot k_{\rm R4}}]} + \\
            &  & +\frac{n({\rm HCO^+})+n({\rm N_2H^+})}{n({\rm H_2})\cdot [1-\frac{n({\rm N_2H^+)}\cdot k_{\rm R9}}{n({\rm N_2})\cdot k_{\rm R4}}]}\;,
    \label{eq:ionisation-luo}
\end{split}
\end{equation}
where $n({\rm M})$ is the volume density of species M.
The volume densities of O and N$_2$ were derived using $x({\rm O})= 2.51 \times 10^{-4}$, and $x({\rm N_2}) = 3.8 \times 10^{-5}$ as in \citet{luo24} and \citet{nielsen26}.
$x{\rm (O)}$ is measured in \citet{lis23} in foreground clouds with H$_2$ volume densities of a few $10^{4}$~\cmc\ toward Sgr B2. 
$x$(N$_2$) is half of the N elemental abundance measured in the local Milky Way by \citet{meyer97}.
To check if the assumed $x({\rm N_2})$ is reasonable in our sources, we have extrapolated the MH$^+$/M trend in Fig.~\ref{fig:affinity} to the proton affinity of N$_2$, which is 493.8 kJ/mole (NIST): we derive that the 
\nnhp/N$_2$ column density ratio should be $\sim 1-3\times 10^{-5}$.
From the measured \nnhp\ column densities in Table~\ref{tab:physical}, we obtain $x({\rm N_2})$ in the range $\sim 1-6\times 10^{-5}$, consistent with the value $3.8\times 10^{-5}$ assumed by \citet{luo24}.
For the other species, we used the abundances measured in this work and converted them to volume densities by multiplying them for $n({\rm H_2})$.

The reaction rates used for Eq.~(\ref{eq:ionisation-luo}) are given in Table~\ref{tab:rates} of Appendix~\ref{app:rates-Luo}, and the resulting ionisation fractions are given in Table~\ref{tab:ionisation-luo}.
We find $x_{\rm L}(e)$ of the order of $10^{-8}$ in the HMSC and HMPO, and of $10^{-7}$ in the UCHII.
Such values are more realistic than those obtained from Eqs.~(\ref{eq:ionisation-cas}) and (\ref{eq:ionisation-noi}), and in line with measurements obtained in similar environments \citep[e.g.,][]{luo24,pineda24,schwarz26}.
Comparing Eqs.~(\ref{eq:ionisation-noi}) and (\ref{eq:ionisation-luo}), part of the reason of such $\sim 3$ orders of magnitude difference could be that $x_{\rm F}(e)$ is dominated, like $x_{\rm C}(e)$, by the factor containing $R_{\rm D}$, not present in Luo's approach.
This factor causes $x_{\rm F}(e)$ to be $\sim 10^{-5}$ when $R_{\rm D}$ is $\sim 10^{-3}$, like in our sources. 
Other differences are the explicit dependence of $x_{\rm L}(e)$ on both $x({\rm N_2H^+})$ and $n({\rm H_2})$, which does not appear in $x_{\rm F}(e)$.
The \citet{luo24} approach computes the ionisation fraction as the sum of the most abundant molecular ions. 
If atomic ions that do not react with H$_2$ constitute a non-negligible fraction of the total ion population, this estimate should be regarded as a lower limit. 
However, chemical models predict that atomic ions become significant in molecular star-forming clouds only under high gas-phase metal abundances \citep{williams98}. 
Such conditions, however, are inconsistent with the well-known sulphur-depletion problem \citep[e.g.,][]{ruffle99}, as substantial depletion of heavy elements is required to reproduce the observed abundances of sulphur-bearing molecules. 
Moreover, following the simplified assumption adopted by \citet{caselli02b}, the depletion of metals can be expected, to first order, to follow that of CO. 
Therefore, in dense regions where CO is heavily depleted, atomic metal ions are also expected to be strongly depleted and should contribute only negligibly to the overall ionisation balance. 
A more quantitative assessment of the role of metal depletion in our targeted regions will require dedicated chemical modelling, which we plan to pursue in future work using models specifically tailored to the physical conditions of our sources.

\subsection{Evolution of the cosmic-ray ionisation rate}
\label{cosmic}

The cosmic-ray ionisation rate can also be estimated following \citet{luo24} using the expression:
\begin{equation}
\zeta_{\rm L} = n({\rm H_2})x({\rm H_3^+})[x({\rm CO})k_{\rm R3} + x({\rm N_2})k_{\rm R4} + x({\rm O})k_{\rm R5} + x_{\rm L}(e)k_{\rm R6}]\;,
    \label{eq:zeta}
\end{equation}
where: $x({\rm CO})$ is derived from the \CII\ or \CIII\ column densities in Table~\ref{tab:neutral}; $x({\rm O})$ and $x({\rm N_2})$ are the same values previously used to compute $x_{\rm L}(e)$;
$k_{\rm R3}$, $k_{\rm R4}$, $k_{\rm R5}$, and $k_{\rm R6}$ are the rates of reactions (R3)-(R6) in Table~\ref{tab:rates};
$x({\rm H_3^+})$ is expressed as:
\begin{equation}
x({\rm H_3^+})=\frac{1}{n({\rm H_2})}\frac{n(e^-)^2-n(e^-)\cdot[n({\rm HCO^+})+n({\rm N_2H^+})]}{n(e^-)+n({\rm O }) k_{\rm R5}/k_{\rm R13}}\;,
    \label{eq:h3p}
\end{equation}
where $k_{\rm R5}$ and $k_{\rm R13}$ are listed in Table~\ref{tab:rates}.
Eq.~(\ref{eq:h3p}) is derived from Eq.~(5)\footnote{note that in \citet{luo24} the numerator of Eq.~(5) contains also the coefficient $\alpha = \frac{n({\rm N_2H^+})n({\rm O})k_{\rm R10}}{k_{\rm R13}}$, which approaches zero considering the updated reaction rates in Table~\ref{tab:rates}.} in \citet{luo24}, which expresses the ${\rm H_3^+}$ volume density.

The values of $\zeta_{\rm L}$ obtained from Eq.~(\ref{eq:zeta}) are listed in Table~\ref{tab:ionisation-luo}.
There is an increase of $\zeta$ in each evolutionary stage, from $\zeta \sim 10^{-16}$~s$^{-1}$ in the HMSC, to $\zeta \sim 10^{-15}$~s$^{-1}$ in the HMPO, to $\zeta \sim 10^{-13}$~s$^{-1}$ in the UCHII region.
The values in the HMPO and UCHII stages, much higher than the typical $\zeta$ adopted for molecular clouds 
\citep[i.e. $\sim 10^{-17}$~s$^{-1}$][]{padovani09,iem12}, 
are consistent with the existence of internal CR sources.
According to model predictions, when protostars (or protoclusters) become active, they accelerate CRs into the densest regions of molecular clouds \citep[e.g.,][]{padovani16,gaches19}, and $\zeta$ can reach values in the range $10^{-16}-10^{-14}$ s$^{-1}$.
As reference, the value measured in the diffuse cloud close to the intermediate-mass protocluster FIR4 in the Orion Molecular Cloud 2 (OMC-2) is $\zeta \sim 3\times 10^{-16}$~s$^{-1}$
\citep{lopez13}, while values up to $4\times 10^{-14}$~s$^{-1}$ are measured in the dense protostellar envelope of FIR4 \citep{ceccarelli14,fontani17}.
\citet{padovani21} indicate a local, embedded source of CRs as the most likely solution for the high $\zeta$ in OMC--2 FIR4.
Similar high values are found in other low- and high-mass protostellar objects \citep[e.g.][]{cabedo23,luo24,schwarz26}.
In particular, \citet{pineda24} noticed an increase in $\zeta$ of orders of magnitudes going from the extended envelope to the dense star forming cores in the nearby star-forming region NGC 1333.
Our findings are consistent with these results, and indicate that both low- and high-mass star-forming cores are associated with $\zeta$ enhancements by orders of magnitude with respect to the canonical ISM value.

The main novelty emerging from our study is that $\zeta$ appears to increase with evolution, particularly from the HMPO to the UCHII stage, possibly due to a progressive temporal strengthening of the internal sources of CRs.
\citet{padovani19} proposed a theoretical mechanism able to accelerate thermal electrons inside HII regions through the first-order Fermi acceleration.
In this scenario, CRs accelerated inside the ionised region could eventually escape to collide with the surrounding gas, and add their effect to those created by protostellar jets still present in the UCHII phase.
However, the very high $\zeta$ towards G589 should also cause a high abundance of PO$^+$ \citep{rivilla22}, which is not apparent from the CHEMICO data.
However, PO$^+$ was observed in a very different source in the galactic center, where the chemistry is regulated by cosmic rays and shocks, thus with a predominance of molecular production in grains. 
In G589, hot core chemistry diluted in the telescope beam could be dominant instead.
Exploring more production channels of PO$^+$, as well as deeper integration observations of G589 in P-bearing species, will allow us to shed light on this seeming paradox.

\section{Summary and conclusions}
\label{conc}

As part of the CHEMICO project, the main objective of this work was to investigate how molecular ions evolve with time during high-mass star formation.
The main results are the following:

(i) We detected 19 molecular ions (including isotopologues) in total in the three CHEMICO targets.
Particularly noteworthy is the first tentative detection in some of the observed targets of the rare species PO$^+$, H$_2$COH$^+$, CH$_3$CO$^+$ and HC$_3$NH$^+$;

(ii) Sulphur-bearing species, in particular SO$^+$, exhibit the clearest evolutionary variations in excitation temperature and linewidths. SO$^+$ is also the unique species showing a clear increase with time of the fractional abundance. 
On the opposite side, nitrogen-bearing ions show weak sensitivity to evolutionary stage, and appear to trace colder and more quiescent gas reservoirs that survive during the entire star-formation process. 
We find a clear decrease in the fractional abundance of the deuterated species DCO$^+$ and, in particular, \nndp, likely caused by the well-established link with the gas temperature enhancement with evolution;

(iii) The protonated-to-neutral abundance ratio increases with the proton affinity of the neutral molecule.
This behaviour was previously observed only in cold dark cloud conditions, and is now demonstrated to remain valid in warm, more evolved stages, suggesting that the proton-transfer process remains similar across a wide range of physical conditions;

(iv) Electron fractions and cosmic-ray ionisation rates, measured from an analytic method tailored for warm gas conditions, increase during evolution, in line with an internal production of cosmic-rays.
In particular, $\zeta$ increases from $\sim 10^{-16}$~s$^{-1}$ in the HMSC stage to $\sim 10^{-15}$~s$^{-1}$ in the HMPO stage, to finally $\sim 10^{-13}$~s$^{-1}$ in the UCHII stage, indicating that the enhancement is particularly efficient in the last stage.

Given the structural complexity of the observed targets, follow-up interferometric observations are essential to delineate the emission morphology of the studied ions.
This will allow us to precisely define the emitting region and, from this, to understand the actual range of the physical parameters associated with the emission of each molecular ion.
High-angular resolution images will also allow us to give better estimate of the source size in each line/tracer, important to give more accurate column density estimates.

\begin{acknowledgements}

This work is based on observations carried out under project number [124-20] and [129-21] with the IRAM 30m telescope. 
IRAM is supported by INSU/CNRS (France), MPG (Germany) and IGN (Spain).
The research leading to these results has received funding from the European Union’s Horizon 2020 research and innovation program under grant agreement No 101004719 [OPR].
We thank O. Sipil\"{a} for useful discussion and checks about the contribution of atomic ions to the electron fraction.
V.M.R., A.L.-G., and L.C. acknowledge support from the grant PID2022-136814NB-I00 by the Spanish
Ministry of Science, Innovation and Universities/State Agency of Research MICIU/AEI/10.13039/501100011033 and by ERDF, UE.
V.M.R. and A.L.-G. also acknowledges support from the grant RYC2020-029387-I funded by MICIU/AEI/10.13039/501100011033 and by "ESF, Investing in your future", and from the Consejo Superior de Investigaciones Cient{\'i}ficas (CSIC) and the Centro de Astrobiolog{\'i}a (CAB) through the project 20225AT015 (Proyectos intramurales especiales del CSIC); and from the grant CNS2023-144464 funded by MICIU/AEI/10.13039/501100011033 and by “European Union NextGenerationEU/PRTR”. 
The project received the support of a fellowship from the “la Caixa” Foundation (ID 100010434). 
The fellowship code is LCF/BQ/PR25/12110012.
H.M-C. acknowledges support from the grant JAE-intro (JAEINT-23-01443) from the CSIC.
This work was supported by the Thematic Action
“Physique et Chimie du Milieu Interstellaire” (PCMI) of INSU Programme National “Astro”, with contributions from CNRS Physique \& Chimie, CEA, and CNES.
\end{acknowledgements}

\onecolumn

\begin{appendix}

\FloatBarrier
\section{Spectra and spectroscopic parameters of the detected lines}
\label{app:A}
\FloatBarrier

\FloatBarrier
In this appendix we show the spectra of the transitions detected, listed in Table~\ref{tab:spectroscopic-tot}.
The non-detections are not shown.
\FloatBarrier

\begin{table*}[h]
\small
    \caption{Spectroscopic parameters of detected lines.}
    \label{tab:spectroscopic-tot}
            \setlength{\tabcolsep}{3pt}
    \begin{tabular}{lccccccc|lcccccc}
            \hline
            \noalign{\smallskip}
 Rest freq. & QN\tablefootmark{a} & $E_{\rm u}$ & Log$_{10}A_{\rm ul}$ & G034 & AFGL & G589 & &      Rest fre. & QN\tablefootmark{a} & $E_{\rm u}$ & Log$_{10}A_{\rm ul}$ & G034 & AFGL & G589 \\
    (MHz)     &  & (K) & (s$^{-1}$) & HMSC & HMPO & UCHII & &   (MHz)     &  & (K) & (s$^{-1}$) & HMSC & HMPO & UCHII \\
            \noalign{\smallskip}
            \hline
            \noalign{\smallskip}
      \multicolumn{7}{c}{NS$^+$} & &   \multicolumn{7}{c}{N$_2$H$^+$} \\ 
 100198.5 & $2-1$& 5.0 & --4.6565 & Y & Y & Y & &   93173.4 & $1-0$  & 4.5 & --4.44034 & Y & Y & Y \\
 150295.6 & $3-2$ & 10.0 & --4.0983 & Y & Y & Y\tablefootmark{d} & &    \multicolumn{7}{c}{N$_2$D$^+$} \\
 250481.5 & $5-4$  & 25.1 & --3.4073 & N & Y & Y\tablefootmark{d} & &   77109.3 & $1-0$  & 3.7 & --4.6869 & Y & Y & N \\
  \multicolumn{7}{c}{PO$^+$\tablefootmark{b}} & &  154217.0 & $2-1$& 11.1 & --3.7047 & Y & Y & Y \\
  94047.8  & $2-1$ & 6.8 & --4.3388 & N & Y & N & &    231321.9 & $3-2$ & 22.2 & --3.1465 & Y & Y & N \\
  \multicolumn{7}{c}{SO$^+$\tablefootmark{b,c}} & &   \multicolumn{7}{c}{$^{15}$NNH$^+$} \\
  115804.4 & $5/2-3/2$ e & 8.9 & --5.1477 & Y? & Y & Y & &   90263.8 & $1-0$  & 4.3 & --4.4817 & N & Y & Y \\
  116179.9 & $5/2-3/2$ f & 8.9 & --5.1435 & Y & Y & Y & &   270783.6 & $3-2$ & 26.0 & --2.9412 & N & Y & Y \\
  162198.6 & $7/2-5/2$ e & 16.7 & --4.6789 & Y? & Y & Y & &   \multicolumn{7}{c}{N$^{15}$NH$^+$} \\
  162574.1 & $7/2-5/2$ f & 16.7 & --4.6757 & Y & Y & Y & &   91205.7 & $1-0$  & 4.4 & --4.4682 & N & Y & Y? \\
  208590.0 & $9/2-7/2$ e & 26.7 & --4.3352 & N & Y & Y & &   273609.0 & $3-2$ & 26.3 & -2.9277 & N & Y & Y? \\
  208965.4 & $9/2-7/2$ f & 26.8 & --4.3329 & N & Y & Y & &   \multicolumn{7}{c}{HCNH$^+$} \\
  254977.9 & $11/2-9/2$ e & 38.9 & --4.0639 & N & Y & Y & &   74111.3 & $1-0$& 3.6 & --6.8768 & N & N & Y \\
  255353.2 & $11/2-9/2$ f & 38.9 & --4.0619 & N & Y & Y & &   148221.5 & $2-1$& 10.7 & --5.8945 & Y & Y & Y \\
  301361.5 & $13/2-11/2$ e & 53.4 & --3.8395 & -- & Y & -- & &   222329.3 & $3-2$ & 21.3 & --5.3363 & Y & Y & Y \\
  301736.8 & $13/2-11/2$ f & 53.5 & --3.8379 & -- & Y & -- & &   \multicolumn{7}{c}{HOCO$^+$\tablefootmark{b}} \\
  \multicolumn{7}{c}{HCO$^+$} & &   85531.5 & $4_{0,4}-3_{0,3}$ & 10.3 & --4.8879 & Y & Y? & Y? \\
  89188.5 & $1-0$  & 4.3 & --4.3781 & Y & Y & Y & &   106913.6 & $5_{0,5}-4_{0,4}$ & 15.4 & --4.5873 & Y & Y & Y \\
 178375.1 & $2-1$& 12.8 & --3.3959 & Y & Y & Y & &   128295.1 & $6_{0,6}-5_{0,5}$ & 21.6 & --4.3432 & Y & Y & Y \\
 267557.6 & $3-2$ & 25.7 & --2.8376 & Y & Y & Y & &  149675.9 & $7_{0,7}-6_{0,6}$ & 28.7 & --4.1376 & Y? & Y & Y \\
 \multicolumn{7}{c}{DCO$^+$}    & &   171056.0 & $8_{0,8}-7_{0,7}$ & 36.9 & --3.9599 & N & Y & Y \\
  144077.3 & $2-1$& 10.4 & --3.1810 & Y & Y & Y & &   213813.4 & $10_{0,10}-9_{0,9}$ & 56.4 & --3.6642 & N & Y & Y \\
  216112.6 & $3-2$ & 20.7 & --2.6228 & Y & Y & Y & &   235190.4 & $11_{0,11}-10_{0,10}$ & 67.7 & --3.5381 & N & Y & Y? \\
  288143.9 & $4-3$ & 34.6 & --2.2322 & -- & Y & -- & &   256566.3 & $12_{0,12}-11_{0,11}$ & 80.0 & --3.4232 & N & Y & N \\
  \multicolumn{7}{c}{H$^{13}$CO$^+$} & &   \multicolumn{7}{c}{H$_2$COH$^+$} \\
  86754.3 & $1-0$  & 4.2 & --4.4142 & Y & Y & Y & &   126923.4 & $2_{0,2}-1_{0,1}$ & 9.1 & --4.7373 & N & Y? & Y \\
  173506.7 & $2-1$& 12.5 & --3.4319 & Y & Y & Y & &   168401.1 & $1_{1,0}-1_{0,1}$ & 11.1 & --4.0572 & N & N & Y \\
  260255.3 & $3-2$ & 25.0 & --2.8737 & Y? & Y & Y & &   173697.5 & $2_{1,1}-2_{0,2}$ & 17.5 & --4.0237 & N & Y?\tablefootmark{d} & N \\
  346998.3 & $4-3$ & 41.6 & --2.4831 & -- & Y & -- & &  226632.6 & $6_{1,5}-6_{0,6}$ & 74.5 & --3.7435 & N & Y\tablefootmark{d} & N \\
   \multicolumn{7}{c}{HC$^{18}$O$^+$}          & &   252870.3 & $4_{0,4}-3_{0,3}$ & 30.4 & --3.7938 & N & Y & N \\ 
  85162.2 & $1-0$  & 4.1 & --4.4383 & Y & Y & Y     & &   \multicolumn{7}{c}{CH$_3$CO$^+$\tablefootmark{e}} \\        
  170322.6 & $2-1$& 12.3 & --3.4561 & Y? & Y & Y    & &   91342.7 & $5_{0}-4_{0}$ & 13.2 & --4.3064 & N & N & Y? \\      
  255479.4 & $3-2$ & 24.5 & --2.8979 & N & Y & Y    & &   91340.9 & $5_{1}-4_{1}$ & 20.0 & --4.3241 & N & N & Y? \\         
  340630.7 & $4-3$ & 40.9 & --2.5073 & -- & Y & --  & &   91335.2 & $5_{2}-4_{2}$ & 40.7 & --4.3822 & N & N & Y\tablefootmark{d} \\
   \multicolumn{7}{c}{HC$^{17}$O$^+$}               & &  164408.9 & $9_{0}-8_{0}$ & 39.5 & --3.5227 & N & N & Y \\ 
  87057.5 & $1-0$  & 4.2 & --4.4096 & N & Y? & Y    & &   164405.5 & $9_{1}-8_{1}$ & 46.3 & --3.5281 & N & N & Y \\     
 174113.2 & $2-1$& 12.5 & --3.4274 & N & Y & Y      & &   164395.4 & $9_{2}-8_{2}$ & 70.0 & --3.5449 & N & N & Y \\    
 261164.9 & $3-2$ & 25.1 & --2.8691 & N & Y & Y     & &   164378.5 & $9_{3}-8_{3}$ & 101.4 & --3.5741 & N & N & Y \\             
  \multicolumn{7}{c}{HCS$^+$}  & &   \multicolumn{7}{c}{HC$_3$NH$^+$\tablefootmark{f}} \\
  85347.9  & $2-1$& 6.1 & --4.9548 & Y & Y & Y & &   86578.1 & $10-9$ & 22.9 & --5.0304 & N & N & Y \\
  128020.6 & $3-2$ & 12.3 & --4.3966 & Y & Y & Y & &   95235.5 & $11-10$ & 27.4 & -4.9044  & N & N & Y \\
  170691.6 & $4-3$ & 20.5 & --4.0060 & Y & Y & Y & &   103892.8 & $12-11$ & 32.4 & --4.7894  & N & N & Y \\
  213360.7 & $5-4$ & 30.7 & --3.7055 & N & Y & Y & &   112549.9 & $13-12$ & 37.8 & -4.6838  & N & N & Y? \\
  256027.1 & $6-5$ & 43.0 & --3.4614 & N & Y & Y & &   129863.8 & $15-14$ & 49.9 & --4.5861  & N & N & Y \\
  298690.5 & $7-6$ & 57.3 & --3.2558 & -- & Y & -- & &   155833.4 & $18-17$ & 71.1 & --4.2554  & N & N & Y \\
  341350.2 & $8-7$ & 73.7 & --3.0782 & -- & Y & -- & &   & & & & & & \\
	 \multicolumn{7}{c}{HC$^{34}$S$^+$} & &   & & & & & & \\
  83965.6 & $2-1$ & 6.0 & --4.9760 & N & N & Y & &   & & & & & & \\
  125947.2 & $3-2$ & 12.1 & --4.4179 & N & Y & Y & &   & & & & & & \\
  167927.3 & $4-3$ & 20.1 & --4.0273 & N & Y & Y?\tablefootmark{d} & &   & & & & & & \\
  209905.3 & $5-4$  & 30.2 & --3.7268 & N & Y? & Y & &   & & & & & & \\
  251880.9 & $6-5$ & 42.3 & --3.4827 & N & Y? & Y?\tablefootmark{d} & &   & & & & & & \\
  \hline
  \end{tabular}
  \tablefoot{Y=detection; N=non detection; Y?=tentative detection.
All parameters are taken from the Cologne Database for Molecular Spectroscopy (CDMS), except when differently specified.
\tablefoottext{a}{Quantum numbers are indicated according to: $J^u-J^l$ for linear molecules, $J^u_{K}-J^l_{K}$ for symmetric-top molecules, $J^u_{K_a,K_b}-J^l_{K_a,K_b}$ for asymmetric-top molecules;}
\tablefoottext{b}{Parameters taken from the Jet Propulsion Laboratory (JPL) catalogue;}
\tablefoottext{c}{All rotational transitions are in the $^2\Pi_{1/2}$ state. The e- and f-parity components are indicated;}
\tablefoottext{d}{Possibly blended;}
\tablefoottext{e}{Parameters taken from \citet{cernicharo21};}
\tablefoottext{f}{The listed Log$_{10}A_{ul}$ is that of the strongest hyperfine component, which is the one with $F$ quantum numbers $F=J+1$ for all lines (for example, for $J^u-J^l=12-11$, it is the one with $F^u-F^l=13-12$).}
    }
\normalsize
\end{table*}

\begin{figure}[h]
\caption{Lines detected towards G034 [HMSC]}
    \includegraphics[width=0.85\linewidth]{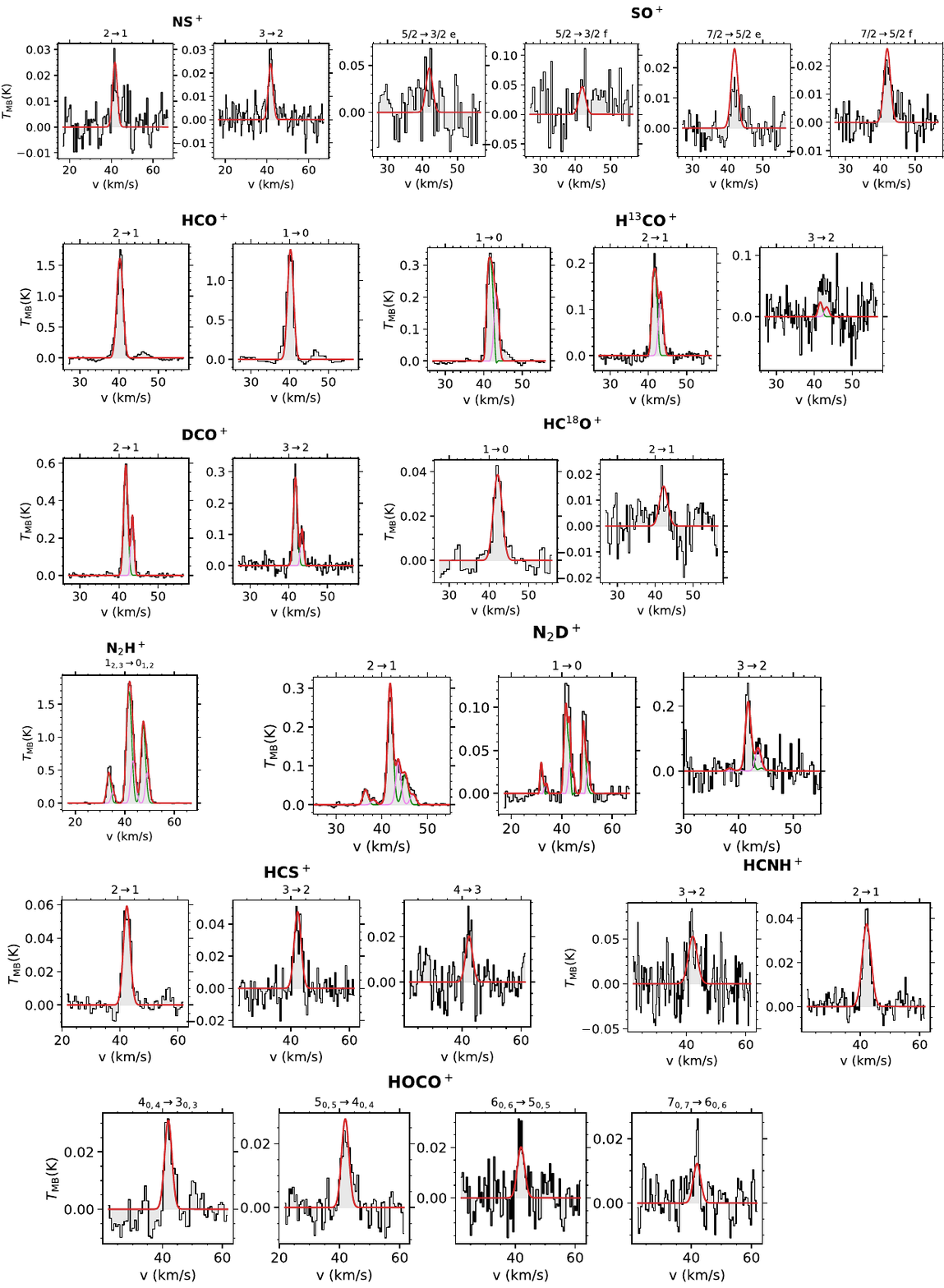}
    \label{fig:spectra-G034}
    \tablefoot{For each species, lines are ordered from left to right from the most to the least intense one.
    Red curves correspond to the best-fit obtained as described in Sect.~\ref{sampleandobs}, taking into account the line eventual hyperfine structure (when resolved).
    The green and magenta curves in the spectra of HCO$^+$, N$_2$H$^+$, and their isotopologues represent the best-fit to the two different velocity features, spectrally well-resolved only in the lines of these species. }
\end{figure}

\begin{figure}[h]
\caption{Lines detected towards AFGL [HMPO]}
    \includegraphics[width=0.85\linewidth]{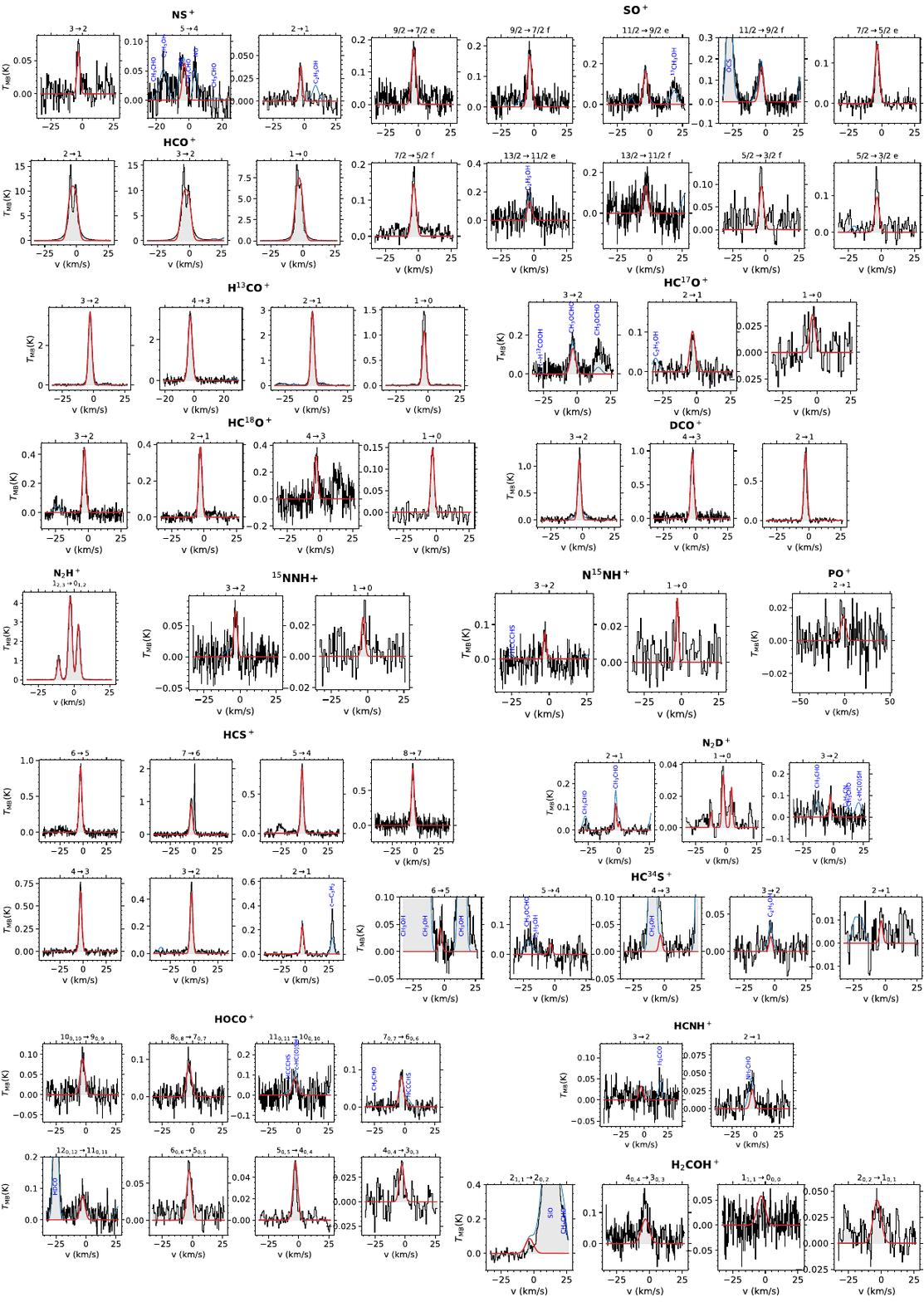}
    \label{fig:spectra-AFGL}
    \tablefoot{Same as Fig.~\ref{fig:spectra-G034} for AFGL. 
    Nearby transitions possibly contaminating the lines of molecular ions are indicated in blue.
    The lines of the main HCO$^+$ isotopologue are not shown.}
\end{figure}

\begin{figure}[h]
\caption{Lines detected towards G589 [UCHII]}
    \includegraphics[width=0.85\linewidth]{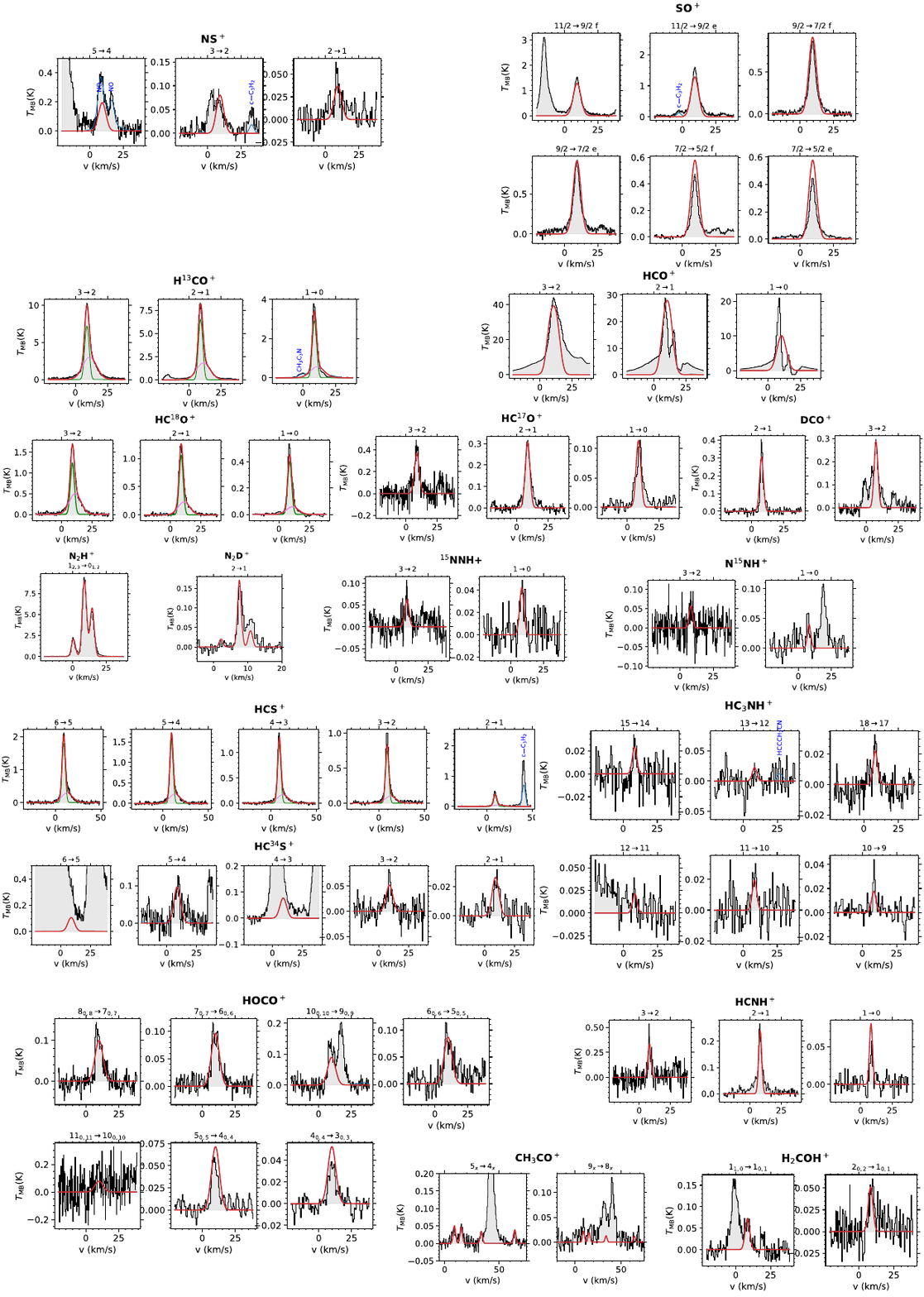}
    \label{fig:spectra-G589}
    \tablefoot{Same as Fig.~\ref{fig:spectra-G034} for G589. 
    Nearby transitions possibly contaminating the lines of molecular ions are indicated in blue.
    The lines of the main HCO$^+$ isotopologue are not shown.}
\end{figure}

\FloatBarrier
\section{Column densities of neutral species}
\label{appNeutral}

\begin{table*}[h]
    \centering
    \caption{Column densities of the neutral species used to compute MH$^+$/M.}
    \label{tab:neutral}
            \setlength{\tabcolsep}{2pt}
    \begin{tabular}{lcccccccccc}
 \noalign{\smallskip}
\hline
\noalign{\smallskip}
Sources & C$^{18}$O  &    C$^{17}$O\tablefootmark{a}   &   H$_2$CO  & H$_2^{13}$CO  &  $^{13}$CS &  C$^{34}$S &   H$^{13}$CN\tablefootmark{a}  &  HN$^{13}$C  & HC$_3$N & H$_2$CCO \\
 \noalign{\smallskip}
     & \multicolumn{10}{c}{(cm$^{-2}$)} \\
 \noalign{\smallskip}
    & $\times 10^{16}$ & $\times 10^{15}$ & $\times 10^{13}$ & $\times 10^{13}$ & $\times 10^{13}$ & $\times 10^{13}$ & $\times 10^{13}$ & $\times 10^{12}$ & $\times 10^{13}$ & $\times 10^{13}$ \\
 \noalign{\smallskip}
\hline
\noalign{\smallskip}
G034[HMSC] & 0.59$\pm$0.05 & 1.4$\pm$0.1 & 0.64$\pm$0.09 & --         & 0.17$\pm$0.12 &  1.0$\pm$0.9 & 1.0$\pm$0.8 &  4.8$\pm$4 &  0.56$\pm$0.06 & -- \\
AFGL[HMPO] & 1.74$\pm$0.02 & 4.68$\pm$0.05 & 38$\pm$1 &  1.07$\pm$0.5 &  1.38$\pm$0.02 & 3.31$\pm$0.04 & 2.05$\pm$0.04  & 3.6$\pm$0.1 &  3.29$\pm$0.06 & 4.6$\pm$0.2\\
G589[UCHII] & 15.6$\pm$0.3 & 59.0$\pm$0.2 & 155$\pm$3 & 8$\pm$2        &  13$\pm$1      & 51$\pm$6 &  13.7$\pm$0.3  &  27$\pm$1  & 35.5$\pm$0.3 & 14.3$\pm$0.9 \\
\noalign{\smallskip}
\hline
\noalign{\smallskip}
\end{tabular}
\tablefoot{Towards G034 and G589, all species were fit considering two velocity components (see description of the origin of the two components in Sect.~\ref{sampleandobs}). 
The listed column densities are the sum of them.
\tablefoottext{a}{Species fit accounting for the line hyperfine structure.}
}
\end{table*}

\FloatBarrier
\section{Derivation of $x(e)$ and $\zeta$ from observations: update of \citet{caselli98}.}
\label{app:opr}

Using the list of chemical reactions in
Table~\ref{tab:chemical-network} we can compute the evolution of the volume density $n$ of each molecular species.
\begin{eqnarray}
 \frac{dn(\ce{H_2^+})}{dt}   &  =  & \zeta \, n(\ce{H_2})    -    k_{2}   \, n(\ce{H_2^+})n(\ce{H_2})   \label{A1}  \\  
\frac{dn(\ce{H_3^+)}}{dt} & =  
                                   & \lbrace  k_2 \, n(\ce{H_2^+}) + [k_{4p}  \frac{1}{1+opr} +k_{4o} \frac{opr}{1+opr}]\,\rbrace \,  n(\ce{H_2}) \, n(\ce{H_2D^+}) - k_3  \, n(\ce{H_3^+})n(\ce{HD})    \label{A2}   \\    \notag
                                      &   & - k_5  \,  n(\ce{H_3^+}) \, n(\ce{CO})    - (k_6 + k_7) \, n(\ce{H_3^+}) \, n(\ce{O}) - (k_{11} + k_{12}) \, n(\ce{H_3^+}) \, n(e^-)  \\
\frac{dn(\ce{H_2D^+)}}{dt} & =  
                                   & k_3 \, n(\ce{H_3^+})n(\ce{HD}) - [k_{4p} \frac{1}{1+opr} + k_{4o} \frac{opr}{1+opr}] \,  n(\ce{H_2D^+}) n(\ce{H_2})  \label{A3} \\ \notag
                                  &   & - (k_{8a} + k_{8b})  \, n(\ce{H_2D^+}) \, n(\ce{CO})  - (k_{9a}+k_{9b} + k_{10a}+k_{10b}) \, n(\ce{H_2D^+})n(\ce{O}) - (k_{13}+k_{14} + k_{15}) \,   n(\ce{H_2D^+}) \, n(e^-)  
\end{eqnarray}
This equation may be simplified by recognising that $k_{8a}+k_{8b} = k_5,  \, k_{9a}+k_{9b} = k_6$ and  $k_{10a}+k_{10b} = k_7$.
Then 
\begin{eqnarray}
\frac{dn(\ce{H_2D^+)}}{dt} & =  
                                   & k_3 \, n(\ce{H_3^+})n(\ce{HD}) - [k_{4p} \frac{1}{1+opr} + k_{4o} \frac{opr}{1+opr}] \,  n(\ce{H_2D^+}) n(\ce{H_2}) \label{A3} \\ \notag
                                  &   & -   [k_5 \, n(\ce{CO}) + (k_{6} + k_{7}) \, n(\ce{O})] \,  n(\ce{H_2D^+}) - (k_{13}+k_{14} + k_{15}) \, n(e^-) \, n(\ce{H_2D^+})
 \end{eqnarray}
 We also introduce $k_{8a} = \frac{2}{3} k_5$ and  $k_{8b} = \frac{1}{3} k_5$ and recognize that the dissociative recombination 
 rates of \ce{HCO+} and \ce{DCO+ } are taken as equal: $ (k_{19a} + k_{19b}+ k_{19c}) =  k_{16} + k_{17} + k_{18}  $
 \begin{eqnarray}
  \frac{dn(\ce{HCO^+})}{dt}   &  =  & k_5 \,   [  n(\ce{H_3^+})+  \frac{2}{3}\,  n( \ce{H_2D^+})]   \, n(\ce{CO}) -  
                                                  (k_{16} + k_{17} + k_{18}) \, n(\ce{HCO^+})n(e^-)          \label{A4}   \\ 
 \frac{dn(\ce{DCO^+})}{dt}   &  =  &    \frac{1}{3}\, k_{5} \, n(\ce{H_2D^+}) n( \ce{CO})  -        (k_{16} + k_{17}+ k_{18}) \,n(e^-) \, n(\ce{DCO^+})   \label{A5}
 \end{eqnarray}

We compute the steady state values $ \frac{dn}{dt} = 0 $  and make the assumption that the abundance of O is close to that of CO, as previously introduced by \citet{caselli98}. 
We may express the electronic fraction $x(e) = \frac{n(e^-)}{n(\ce{H_2})}$ and the cosmic ionisation rate per \hh, $\zeta$, in terms of the observationally derived quantities  
$\ce{R_D} = \frac{\ce{DCO^+}}{\ce{HCO+}} $ and $\ce{R_H} = \frac{\ce{HCO^+}}{\ce{CO}}= \frac{\ce{H^{13}CO^+}}{\ce{^{13}CO}}$.
We derive R$_D$ from reactions \ref{A4} and \ref{A5}: 
\begin{equation}
   n(\ce{HCO+)} =  \frac{ k_5 \, [ n(\ce{ H_3^+}) + \frac{2}{3}\,  n(\ce{ H_2D^+)] } }{ (k_{16} + k_{17}+k_{18}) \, n(e^-) }  n(\ce{CO})  
   \label{A6}
\end{equation}
\begin{equation}
   n(\ce{DCO+)} =  \frac{ k_5 \frac{1}{3} \, n(\ce{ H_2D^+) } }{(k_{16} + k_{17}+k_{18}) \, n(e^-)  }n \ce{(CO)}  
   \label{A7}
\end{equation}
\begin{equation}
   \ce{R_D}   =     \frac{1}{3}  \cdot \frac{ n(\ce{ H_2D^+) }} {  n(\ce{H_3^+)} + \frac{2}{3} \,  n(\ce{H_2D^+) } }  
   \label{A8}
\end{equation}
 and relation \ref{A6} allows to express $\ce{R_H}$
   \begin{eqnarray}
     \ce{R_H}  & = &   \frac{ k_5 \, [n(\ce{ H_3^+}) + \frac{2}{3} \ n(\ce{ H_2D^+)] } }{ (k_{16} + k_{17}+k_{18}) \, n(e^-) }  
 \end{eqnarray}
 The ratio $ \frac{ n(\ce{ H_2D^+) }} {  n(\ce{H_3^+)} }$ is directly derived from Equation \ref{A3}. We define $k_e$ as the total dissociative recombination rate of \hhdp.
 \begin{equation}
 \begin{split}
 \frac{ n(\ce{ H_2D^+) }} {  n(\ce{H_3^+)} }& = &\frac{k_3 \, n(\ce{HD})} { [ \frac{1}{1+opr} k_{4p} + \frac{opr}{1+opr} k_{4o} ]\ n(\ce{H_2}) + (k_{5} + k_{6}+k_{7}) \ n(\ce{CO}) + k_e \ n(e^-)}  \\
                             &  =    &    \frac{k_3  \, x(\ce{HD})}{ [k_{4p}  \frac{1}{(1+opr)} +k_{4o} \frac{opr}{(1+opr)}]  + (k_5+k_6+k_7) \, x(\ce{CO}) + 
                             k_e x(e)} = \frac{3 \ce{R_D}}{1 - 2 \ce{R_D}}  \\
 \label{A11}
 \end{split}
 \end{equation} 
This expression accounts for the possible reverse endothermic reaction between \hh\ and \hhdp. 
We recall that $k_{4p} = 2.0\times 10^{-9} \, \rm{exp}(-232/T) $ and  $k_{4o} = 2.0\times 10^{-9} \, \rm{exp}(-61.5/T) $.
Replacing the reaction rate constants by their values displayed in Table~\ref{tab:chemical-network} and taking $x(\ce{HD})  = 3  \times 10^{-5}$, we obtain:
\begin{equation}
 \frac{ n(\ce{ H_2D^+) }} {  n(\ce{H_3^+)} }   =    \frac{6 \times 10^{-14}}{2.0\cdot 10^{-9} [\frac{ \rm{exp}(-232/T)}{1+opr} +  \frac{opr \cdot \rm{exp}(-61.5/T)}{1+opr}]+ [1.7 \cdot 10^{-9}+1.75 \cdot 10^{-9} (T/20)^{-0.16}{ \rm{exp}}(-1.4/T)]   x(\ce{CO})+ 3.48 \cdot 10^{-7} (T/20)^{-0.65}   x(e)} \label{A12}
 \end{equation}
The electronic fraction $x(e)$ can be derived from equations \ref{A11} and \ref{A12} where $x(\ce{CO}) $ is expressed as $2.7  \times 10^{-4} / f_D$
\begin{eqnarray}
x(e) &= & ( \frac{T}{20} ) ^{0.65} \times \,\,  \Biggl[ \frac{5.7\times 10^{-8}}{\ce{R_D}}-1.15 \times 10^{-7} - 5.7 \times 10^{-3}\, \left( \frac{{\rm{exp}}(-232/T)}{1+opr} 
                            + \frac{opr \cdot {\rm{exp}}(-61.5/T)}{1+opr} \right)   \notag \\  
  & - &  \frac{4.2\times 10^{-7}\times {\rm{exp}}(-1.4/T)\times (T/20)^{-0.16}+4.1 \times 10^{-7}}{{f_D}} \Biggr ] \\ \notag
\end{eqnarray}
 \hhhp ~ is formed through reaction \ref{A2} where $n(\ce{H_2+)}$ is directly obtained from the formation/destruction balance described by reactions \ref{A1} and \ref{A2}  as  $  n(\ce{H_2+)}  = \frac{\zeta } {k_2}$.
We can now derive $\zeta$ when computing x(\hhhp) with the same hypotheses.
 \begin{eqnarray*}
  x(\ce{H_3^+})& =   &(1-2\ce{R_D}) \times \ce{R_H}\times \frac{k_{16}+k_{17}+k_{18}}{k_5}x(e)     \notag \\
     & = &  \frac{\zeta}{n(\ce{H_2})}\times \frac{1}{ [k_{4p} \frac{1}{1+opr} + k_{4o} \frac{opr}{1+opr}] + k_3\, x(\ce{HD)}+(k_5+k_6+k_7) \, x(\ce{CO})+(k_{11}+k_{12})\, x(e)}  
 \end{eqnarray*}

 %
 %
We finally get  : 
\begin{eqnarray}
\frac{\zeta}{n(\ce{H_2})} &=& \ce{R_H (1-2R_D)}\,x(e)  \times \Biggl[ \Biggl( 6\times10^{-11}
+ 2 \times10^{-6} \left( \frac{\exp(-232.6/T)}{1+opr}
+ \frac{opr\,\exp(-61.5/T)}{1+opr} \right)  \notag  \\
     &  + &  \frac{0.8\times10^{-10} \left[1.7+1.75 (\frac{T}{20})^{-0.16} \exp(-1.4/T) \right]}{f_D} \Biggr) 
     \left(\frac{T}{20} \right)^{-0.79}  + 2.54\times10^{-4}\,x(e)\,\left(\frac{T}{20}\right)^{-1.31} \Biggr]
 \end{eqnarray}

   \begin{table*}
     \begin{center}
       \caption{Updated chemical network considered for deriving the equations in Sect.~\ref{app:opr}.} 
       \label{tab:chemical-network}
       \begin{tabular}{lccc}\toprule
         Chemical reaction & Rate & Rate value & Reference \\
     \midrule
         \ce{H_2 + CRP -> H_2^+ + e^-}   & $k_1$   & $9.60 \times 10^{-1} \zeta_{H_2}$ & (1)  \\
         \ce{H_2^+ + H_2 -> H_3^+ + H}   & $k_2$   & $2.08 \times 10^{-9}$ & (2) \\
         \ce{H_3^+ + HD -> H_2D^+ + H_2} & $k_3$   & $2.00 \times 10^{-9}$ &  (3) \\
         \ce{H_2D^+ + H_2 -> H_3^+ + HD} & $k_4$   & $2.00 \times 10^{-9} \exp(-232.0/T)$ & (3)*\\
         \ce{H_2D^+ + H_2(para) -> H_3^+ + HD} & $k_{4p}$   & $2.00 \times 10^{-9} \exp(-232.0/T)$ &  \\
         \ce{H_2D^+ + H_2 (ortho) -> H_3^+ + HD} & $k_{4o}$   & $2.00 \times 10^{-9} \exp(-61.5/T)$ &  \\
         \ce{H_3^+ + CO -> HCO^+ + H_2}  & $k_5$   & $1.70 \times 10^{-9}$ & (4)\\
         \ce{H_3^+ + O -> OH^+ + H_2}    & $k_6$   &  $1.23 \times 10^{-9}$ (T/20)$^{-0.16}$ exp(-1.4/$T$) & (2)\\
         \ce{H_3^+ + O -> H_2O^+ + H}    & $k_7$   & $5.24  \times 10^{-10}$ (T/20)$^{-0.16}$ exp(-1.4/$T$)  & (2)\\
         \ce{H_2D^+ + CO -> HCO^+ + HD}  & $k_{8a}=\frac{2}{3}k_{5}$ & $1.13 \times 10^{-9}$ & $\dag$ \\
         \ce{H_2D^+ + CO -> DCO^+ + H_2} & $k_{8b}=\frac{1}{3}k_{5}$ & $5.67 \times 10^{-10}$ &$\dag$ \\
         \ce{H_2D^+ + O -> OH^+ + HD}   & $ k_{9a}=\frac{2}{3}k_6$   & $8.21 \times 10^{-10}$  (T/20)$^{-0.16}$ exp(-1.4/$T$) & $\dag$ \\
         \ce{H_2D^+ + O -> OD^+ + H_2}    & $k_{9b}=\frac{1}{3}k_6$   & $4.11 \times 10^{-10}$ (T/20)$^{-0.16}$ exp(-1.4/$T$)& $\dag$ \\
          \ce{H_2D^+ + O -> HDO^+ + H}    & $k_{10a}=k_7/2$   & $2.62\times 10^{-10}$ (T/20)$^{-0.16}$ exp(-1.4/$T$) & $\dag$ \\
         \ce{H_2D^+ + O -> H2O^+ + D}    & $k_{10b}=k_7/2$   & $2.62 \times 10^{-10}$ (T/20)$^{-0.16}$ exp(-1.4/$T$) &$\dag$ \\
      \ce{H_3^+ + e^- -> H + H + H}   & $k_{11}$ & $1.78 \times 10^{-7} (T/20)^{-0.52}$ & (2) \\
         \ce{H_3^+ + e^- -> H_2 + H}     & $k_{12}$ & $9.57 \times 10^{-8} (T/20)^{-0.52}$ & (2) \\
         \ce{H_2D^+ + e^- -> H_2 + D}   & $k_{13}$ & $2.44 \times 10^{-8} (T/20)^{-0.65}$ & (5)\\
        \ce{H_2D^+ + e^- -> HD + H}   & $k_{14}$ & $7.0 \times 10^{-8} (T/20)^{-0.65}$ & (5)\\
         \ce{H_2D^+ + e^- -> H + H + D}  & $k_{15}$ & $2.54 \times 10^{-7} (T/20)^{-0.65}$ & (5)\\
         \ce{H^{13}CO^+ + e^- -> ^{13}CO + H}      & $k_{16}$ & $1.48 \times 10^{-6} (T/20)^{-0.79}$ &  (6)\\
         \ce{H^{13}CO^+ + e^- -> ^{13}C + OH}      & $k_{17}$ & $1.53 \times 10^{-7} (T/20)^{-0.79}$ & (6)\\
         \ce{H^{13}CO^+ + e^- -> O + ^{13}CH}      & $k_{18}$ & $6.8 \times 10^{-8} (T/20)^{-0.79}$ & (6)\\
         \ce{DCO^+ + e^- -> CO + D}      & $k_{19a}=k_{16}$ & $1.48 \times 10^{-6} (T/20)^{-0.79}$ &$\dag$ \\
         \ce{DCO^+ + e^- -> OD + C}      & $k_{19b}=k_{17}$ & $1.53 \times 10^{-7} (T/20)^{-0.79}$& $\dag$ \\
         \ce{DCO^+ + e^- -> O  + CD}      & $k_{19c}=k_{18}$ & $6.8\times 10^{-8} (T/20)^{-0.79}$& $\dag$ \\
         \bottomrule
       \end{tabular}
     \end{center}
   {\small{ (1) A. Dalgarno (pers. comm.); (2) KIDA database, (3) \citet{hugo09}; * assuming \hh ~in para form only; (4) \citet{anicich93};
  (5) Larsson (pers. comm., reported in Caselli et al.~\citeyear{caselli98}); (6) \citet{hamberg14}. Same rate assumed for \ce{HCO^+}; $\dag$ same total rate than for non deuterated species. }}
   \end{table*}

\FloatBarrier
\section{Reaction rates used in the derivation of $x_{\rm L}(e)$ and $\zeta_{\rm L}$.}
\label{app:rates-Luo}

\begin{table}[h]
    \centering
    \caption{Reaction rates used in Eqs.(\ref{eq:ionisation-luo}) and (\ref{eq:zeta}).}
    \setlength{\tabcolsep}{2pt}
    \begin{tabular}{lcc}
\hline
\noalign{\smallskip}
      Chemical reaction   & Rate & Rate value \\
\noalign{\smallskip}
\hline
\noalign{\smallskip}
    \ce{H_3^+ + CO} $\rightarrow$ \ce{HCO^+ + H_2}   & $k_{\rm R3a}$ & $1.70\times 10^{-9}$\tablefootmark{a} \\
    \ce{H_3^+ + CO} $\rightarrow$ \ce{HOC^+ + H_2}   & $k_{\rm R3b}$ & $1.0\times 10^{-10}$ \\
    \ce{H_3^+ + N_2} $\rightarrow$ \ce{N_2H^+ + H_2} & $k_{\rm R4}$ & $1.7\times 10^{-9}$ \\
    \ce{H_3^+ + O} $\rightarrow$ \ce{OH^+ + H_2} & $k_{\rm R5a}$ & $1.23\times 10^{-9}(\frac{T}{20})^{-0.16}{\rm exp}(-1.4/T)$ \\
    \ce{H_3^+ + O} $\rightarrow$ \ce{H_2O^+ + H_2} & $k_{\rm R5b}$ & $5.24\times 10^{-10}(\frac{T}{20})^{-0.16}{\rm exp}(-1.4/T)$  \\
    \ce{H_3^+ + e^-} $\rightarrow$ \ce{H_2 + H} & $k_{\rm R6a}$  & $1.78\times 10^{-7}(\frac{T}{20})^{-0.52}$  \\
    \ce{H_3^+ + e^-} $\rightarrow$ \ce{H + H + H} & $k_{\rm R6b}$  & $9.57\times 10^{-8}(\frac{T}{20})^{-0.52}$  \\
    \ce{N_2H^+ + CO} $\rightarrow$ \ce{HCO^+ + N_2} & $k_{\rm R8}$  & $8.8\times 10^{-10}$  \\
    \ce{N_2H^+ + e^-} $\rightarrow$ \ce{H + N_2} & $k_{\rm R9}$  & $2.40\times 10^{-6}(\frac{T}{20})^{-0.84}$  \\
    \ce{N_2H^+ + O} $\rightarrow$ \ce{OH^+ + N_2} & $k_{\rm R10}$  & $1.4\times 10^{-10}{\rm exp}(-3400/T)$ \\
    \ce{H_3O^+ + e^-} $\rightarrow$ \ce{OH + H_2} & $k_{\rm R13a}$  &  $2.32\times 10^{-7}(\frac{T}{20})^{-0.5}$ \\
    \ce{H_3O^+ + e^-} $\rightarrow$ \ce{OH + H + H} & $k_{\rm R13b}$  &  $1.01\times 10^{-6}(\frac{T}{20})^{-0.5}$ \\
    \ce{H_3O^+ + e^-} $\rightarrow$ \ce{H + H_2O} & $k_{\rm R13c}$  &  $4.26\times 10^{-7}(\frac{T}{20})^{-0.5}$ \\
\noalign{\smallskip}
\hline
    \end{tabular}
    \tablefoot{All rate values were taken from the KIDA database except when differently specified.
    The total rate for a reaction with multiple channels, indicated with subscripts ${a}$, ${b}$, etc., is the sum of the rates of all individual channels.
    For rates depending on temperature, we used \Tk\ given in Table~\ref{tab:sources}
    \tablefoottext{a}{\citet{anicich93};}
    }
    \label{tab:rates}
\end{table}


%
%

\end{appendix}

\end{document}